\documentclass[useAMS,usenatbib]{mn2e}
\usepackage{subfigure}
\usepackage{graphicx}
\usepackage{times}
\usepackage{tabularx}
\usepackage{natbib}
\usepackage{url} 
\usepackage{xcolor}
\usepackage{caption}
\usepackage{amsmath}
\usepackage{float}
\usepackage{multirow}
\usepackage{morefloats}
\usepackage{amssymb}
\usepackage[breaklinks,colorlinks,citecolor=black]{hyperref}

\newcommand{\apss}{APSS}
\newcommand{\apjs}{ApJS}
\newcommand{\apj}{ApJ}

\newcommand{\apjl}{ApJ Letters}
\newcommand{\mnras}{MNRAS}
\newcommand{\aap}{A\&A}

\newcommand{\ssr}{Science \& Space Review}

\newcommand{\nat}{Nature}
\newcommand{\physrep}{Physics Reports}

\newcommand{\araa}{Annual Review of Astronomy and Astrophysics}

\newcommand{\CRaTer}{{\texttt {CRaTer}}}
\newcommand{\flash}{{\texttt {Flash}}}
\newcommand{\dd}{\mathrm{d}}
\newcommand{\Mpc}{\mathrm{Mpc}}
\newcommand{\Msun}{\mathrm{M}_{\odot}}

\newcommand{\kpc}{\mathrm{kpc}}

\newcommand{\K}{\mathrm{K}}
\newcommand{\Gyr}{\mathrm{Gyr}}
\newcommand{\Myr}{\mathrm{Myr}}

\newcommand{\fc}{F_{\mathrm{comp}}}
\newcommand{\feff}{F_{\mathrm{eff}}}

\newcommand{\fst}{F_{\mathrm{stretch}}}
\newcommand{\fb}{F_{\mathrm{baro}}}
\newcommand{\fdiss}{F_{\mathrm{diss}}}
\newcommand{\fadv}{F_{\mathrm{adv}}}

\newcommand{\vvec}{\mathbf{v}}
\newcommand{\pc}{\mathrm{pc}}

\definecolor{myred}{rgb}{1,0,0} 
\definecolor{myblue}{rgb}{0,0,1}
\definecolor{mygreen}{rgb}{0,1,0}

\def\stacksymbols #1#2#3#4{\def\theguybelow{#2}
        \def\verticalposition{\lower#3pt}
        \def\spacingwithinsymbol{\baselineskip0pt\lineskip#4pt}
        \mathrel{\mathpalette\intermediary#1}}
\def\intermediary #1#2{\verticalposition\vbox{\spacingwithinsymbol
        \everycr={}\tabskip0pt
        \halign{$\mathsurround0pt#1\hfil##\hfil$\crcr#2\crcr
                \theguybelow\crcr}}}

\def\lta{\stacksymbols{<}{\sim}{2.5}{.2}}
\def\gta{\stacksymbols{>}{\sim}{2.5}{.2}}

\begin{document}
 
 \title[The AGN feedback chaotic weather]{Dissecting\,the\,turbulent\,weather\,driven\,by\,mechanical\,AGN\,feedback}
 \author[Wittor \& Gaspari]{D. Wittor$^{1,2,3}$\thanks{%
 E-mail: denis.wittor@unibo.it} \& M. Gaspari$^{4,5}$\\
 $^{1}$ Hamburger Sternwarte, Gojenbergsweg 112, 21029 Hamburg, Germany \\
 $^{2}$ Dipartimento di Fisica e Astronomia, Universita di Bologna, Via Gobetti 93/2, I-40122 Bologna, Italy \\
 $^{3}$ INAF - Istituto di Radioastronomia di Bologna, via Gobetti 101, I-41029 Bologna, Italy \\
 $^{4}$ Department of Astrophysical Sciences, Princeton University, 4 Ivy Lane, Princeton, NJ 08544-1001, USA \\
 $^{5}$ INAF - Osservatorio di Astrofisica e Scienza dello Spazio, via P. Gobetti 93/3, I-40129 Bologna, Italy \\
  \vspace{-1.0cm}
  }
 \date{Accepted ???. Received ???; in original form ???}
 \maketitle

 \begin{abstract}
  Turbulence in the intracluster, intragroup, and circumgalactic medium plays a crucial role in the self-regulated feeding and feedback loop of  central supermassive black holes. We dissect the three-dimensional turbulent `weather' in a high-resolution Eulerian simulation of active galactic nucleus (AGN) feedback, shown to be consistent with multiple multi-wavelength observables of massive galaxies. 
  We carry out post-processing simulations of Lagrangian tracers to track the evolution of enstrophy, a proxy of turbulence, and its related sinks and sources. This allows us to isolate in depth the physical processes that determine the evolution of turbulence during the recurring strong and weak AGN feedback events, which repeat self-similarly over the Gyr evolution. 
  We find that the evolution of enstrophy/turbulence in the gaseous halo is highly dynamic and variable over small temporal and spatial scales, similar to the chaotic weather processes on Earth. We observe major correlations between the enstrophy amplification and recurrent AGN activity, especially via its kinetic power. 
  While advective and baroclinc motions are always sub-dominant, \textit{stretching} motions are the key sources of the amplification of enstrophy, in particular along the jet/cocoon, while \textit{rarefactions} decrease it throughout the bulk of the volume. This natural self-regulation is able to preserve, as ensemble, the typically-observed subsonic turbulence during cosmic time, superposed by recurrent spikes via impulsive anisotropic AGN features (wide outflows, bubbles, cocoon shocks).
  This study facilitates the preparation and interpretation of the thermo-kinematical observations enabled by new revolutionary X-ray IFU telescopes, such as \textit{XRISM} and \textit{Athena}.
 \end{abstract}
 \label{firstpage}
 \begin{keywords}
 galaxies: clusters: intracluster medium - hydrodynamics - turbulence - (galaxies:) quasars: supermassive black holes - galaxies: active - methods: numerical
  \vspace{-0.5cm}
 \end{keywords}
 
 \section{Introduction}\label{sec::intro}
 The intracluster medium (ICM) forms the gaseous atmospheres filling galaxy clusters: massive objects with masses of $\sim$\,$10^{15}\ \Msun$ that consist of dark matter ($\sim$\,$85\%$), gas ($\sim$\,$13\%$) and stars ($\sim$\,$2\%$). Similar to Earth weather, the ICM is a highly complex hydrodynamical system that is continuously disturbed by turbulence, shock waves and galaxy motions. During the processes of hierarchical structure formation, the ICM is shock heated to temperatures of $10^7-10^8 \ \K$ \citep[][]{2012_Kravtsov_Borgani} and then cools due to plasma X-ray radiative emission. However, the feedback from the central supermassive black hole (SMBH), a.k.a. active galactic nucleus (AGN), will prevent the runaway of a pure cooling flow catastrophe towards the core of the galaxy cluster or group \citep[e.g.][]{Peterson:2006,2007_McNamara_Nulsen, Fabian:2012}. \\
 \indent
 Such AGN feedback response is tightly linked to the feeding of the central SMBH in a self-regulated cycle. Throughout the past decade, observations and simulations have shown that SMBHs in massive galaxies grow mainly via Chaotic Cold Accretion (CCA), which is best understood as raining onto black holes: here, 
 warm filaments and cold clouds condense out of the turbulent atmosphere and rain onto the SMBH at the center of each galaxy, in particular of the most massive galaxy of the cluster/group halo \citep[][]{Gaspari:2013_cca, Gaspari:2017_cca, Gaspari:2019, Voit:2015_nat,Voit:2015_gE, Voit:2018, Barai:2016, Prasad:2017, McDonald:2018, 2018ApJ...865...13T,2018ApJ...858...17T, Juranova:2019, Rose:2019, Storchi-Bergmann:2019}.
 In response, the AGN feedback is triggered, when the binding energy of the infalling and inelastically colliding clouds is converted into mechanical energy. This mechanical energy is released into the ambient medium predominantly via collimated jets and ultrafast outflows \citep[e.g.][]{2004A&A...413..535G, Tombesi:2013, Sadowski:2017}.
 On larger scales, $\sim5-100\ {\rm kpc}$, AGN feedback drives turbulence, shocks, and bubbles \citep[e.g.][]{Brighenti:2003, gitti:2012, 2012_McNamara_Nulsen, HL:2015, Gaspari:2017_uni, Hillel:2017, Liu:2019, Yang:2019}. Thus, AGN heating quenches the SMBH feeding/cooling for several $\Myr$, before the feeding process re-starts. Hence, the feeding and feedback processes are linked through a self-regulated loop that acts over a large range of scales, i.e. from $\mathrm{mpc}$ to $\Mpc$ and vice versa  (\citealt{Gaspari:2020}, for a review). \\
 \indent
 To advance our understanding of the above AGN feeding/feedback cycle, it is crucial to understand the origin and evolution of the hot halo turbulence, which can alter the heating, cooling, and transport processes over the long-term evolution of the gaseous halo.
 The \textit{Hitomi} X-ray telescope has probed directly such turbulent motions in the Perseus cluster, finding a line-of-sight velocity dispersion of $\approx164\ \rm km\,s^{-1}$ \citep{2016Natur.535..117H}. At large cluster radii, $\sim \Mpc$, mergers and galaxy motions are mainly responsible for the turbulent motions \citep[e.g.][]{2011A&A...529A..17V,2015ApJ...800...60M,2017MNRAS.471.3212W}. On the other hand, AGN feedback is the main driver of turbulence inside the cluster core, $r < 100 \ \kpc$ \citep[e.g.][for a review]{Vazza:2012, Yang:2016a, Lau:2017, Simionescu:2019}. Hence, understanding turbulence linked to the AGN jet/outflows is one of the milestones in studying the self-regulated loop of AGN feedback and feeding. With the advent of groundbreaking Integral-Field-Unit (IFU) X-ray spectrometers (e.g. {\it XRISM} and {\it Athena}), which will be able to unveil turbulent motions down to kpc scales, it is now even more pressing to explore the detailed properties of AGN turbulence.\\
 \indent
 
 To achieve the above goals, in this contribution (which is part of the \textit{BlackHoleWeather} program; \citealt{Gaspari:2020}), we combine state-of-the-art Eulerian and Lagrangian simulations to study turbulence injected by the AGN, in particular in mechanical form -- the long-term maintenance mode of feedback. We dissect the weather driven by the AGN feedback, focusing on the evolution of \textit{enstrophy}, the primary tracer of solenoidal turbulence, and all its related source and sink terms. In Sec. \ref{sec::ens}, we provide an overview of enstrophy and its related physics. In Sec. \ref{sec::simulations}, we introduce our Eulerian and Lagrangian simulations. We present and discuss the results of the Eulerian and Lagrangian analysis in Sec. \ref{sec::results_grid} and \ref{sec::results_tracers}, respectively. The summary and conclusions are given in Sec. \ref{sec::conclusion}.
\subsection{Enstrophy and physical sources} \label{sec::ens}
 \indent
 Enstrophy, the magnitude of vorticity $\epsilon = \frac{1}{2} |\boldsymbol{\nabla} \times \vvec|^2$, is a primary proxy of solenoidal turbulence \citep[e.g.][]{2015ApJ...810...93P,2017MNRAS.464...210V,2017MNRAS.471.3212W}. 
 As enstrophy is a measure for the solenoidal component of the turbulent velocity field, it can be understood as the kinetic energy per unit mass and area of the hydrodynamical flow (with units of [time$^{-2}$]).
 In other words, it can be seen as the squared magnitude of the frequency of the ensemble vortex tubes. In subsonic turbulence (like the ICM here), most of the motions are solenoidal, thus enstrophy provides an excellent measure of the relative magnitude of the driven turbulence.
 Thus, it is related to the dissipation rate of the specific turbulent energy and to the fraction/number of turbulent eddy turnovers per characteristic injection time\footnote{Here of the order of an AGN bubble size, $L\sim 10\ \kpc$.}.

 In the Eulerian frame, the evolution of enstrophy is governed by compressive, baroclinic and stretching motions, as well as by advection and dissipation \citep[e.g.][]{2015ApJ...810...93P}. The evolution of enstrophy in the Lagrangian frame is derived from the Eulerian frame \citep[see][]{2017MNRAS.471.3212W}. For pure hydrodynamics, the equations that describe the evolution of enstrophy take the form:
 \indent
 \begin{align}
 &\left(\frac{\dd \epsilon}{\dd t}\right)_{\mathrm{Euler}} = \fadv + \fc + \fst + \fb + \fdiss. \label{eq::enst_euler} \\
 &\left(\frac{\dd \epsilon}{\dd t}\right)_{\mathrm{Lagrange}} = 2 \fc + \fst + \fb + \fdiss. \label{eq::enst_lagrange}
 \end{align}
 \indent
 The dynamical terms on the right-hand-sides (RHS) of Eq. \ref{eq::enst_euler} and \ref{eq::enst_lagrange} account for different physical processes that generate, amplify, and weaken enstrophy. The conservative advection of enstrophy is described by the advection term $\fadv$. In the Lagrangian frame, the advection is incorporated into both the time derivative and the factor $2$ in front of the compression term. The compression term, $\fc$, describes both the enstrophy amplification by compression and its reduction by rarefaction. We note that, in the Lagrangian analysis (Sec. \ref{sec::results_tracers}), we will refer to $2\fc$ when we analyze the compression term. The amplification of enstrophy through vortex stretching is described by the stretching term $\fst$. The baroclinic term, $\fb$, accounts for the enstrophy generation in non-barotropic and stratified atmospheres. For an adiabatic equation of state, as in our case, this corresponds to the development of enstrophy due to unaligned gradients of density and pressure. Such a scenario might occur behind curved or interacting major shocks, as well as in regions of strong radiative cooling inside a cooling flow.
 The viscous dissipation of solendoidal flows is described by the dissipation term $\fdiss$, that is mainly dominated by the damping of turbulent eddies. Specifically, the different dynamical terms are computed as follows:
 \indent
 \begin{align}
  \fadv &= - \boldsymbol{\nabla} \cdot (\vvec \epsilon) \label{eq:Fadv} \\
  \fc &= -\epsilon \boldsymbol{\nabla}  \cdot \vvec  \label{eq:Fcomp}\\
  \fst &= 2\epsilon (\boldsymbol{\hat{\omega}} \cdot \boldsymbol{\nabla}  ) \vvec \cdot \boldsymbol{\hat{\omega}}  \label{eq:Fstretch}\\
  \fb &= \frac{\boldsymbol{\omega}}{\rho^2} \cdot ( \boldsymbol{\nabla}  \rho \times \boldsymbol{\nabla}  P)  \label{eq:Fbaro}\\
  \fdiss &= \nu \boldsymbol{\omega} \cdot \left({\nabla}^2 \boldsymbol{\omega} + \boldsymbol{\nabla}  \times \mathbf{G}\right) \label{eq:Fdiss} \\
  \mathrm{with} \ \ \boldsymbol{\omega}  &= \boldsymbol{\nabla}  \times \vvec \ \mathrm{and} \ \mathbf{G} = (1/\rho) \boldsymbol{\nabla}  \rho \cdot \mathbf{S}, 
  \label{eq:curl}
 \end{align}
 \noindent
where $P$, $\rho$ and $\nu$ are the pressure, density and kinematic viscosity, respectively, and $\mathbf{S}$ is the traceless strain tensor \citep{2006MNRAS.370..415M}. We define the effective dynamical term, $\feff$ (the net growth/decay rate),  as the sum of all the terms (positive and negative) on the RHS of Eq. \ref{eq::enst_euler} and \ref{eq::enst_lagrange}. Since we have no physical viscosity in our simulation and numerical viscosity is low, we neglect $\fdiss$. We compute the relative contribution of each dynamical term to the effective term as:
 \indent
 \begin{align}
   F_{{\rm rel}, i}= \left|\frac{F_i}{\feff}\right| \label{eq::frel}
 \end{align}
Here, $i$ accounts for the different types of motions, Eq. \ref{eq:Fadv}-\ref{eq:Fbaro}. As an additional observable, we use the effective growth time of enstrophy:
 \begin{align}
   t_{\mathrm{growth}} (t) = \left| \frac{\epsilon(t)}{F_{\mathrm{eff}}} \right|. \label{eq::turntime}
 \end{align}
 Being an unsigned quantity, it does not indicate whether enstrophy is growing or decaying. The effective timescale is correlated with the turbulence turnover rate.
 \section{Numerical simulations}\label{sec::simulations}
 We use both Eulerian and Lagrangian numerical simulations of self-regulated AGN feeding and feedback in a typical massive galaxy. In this Section, we summarize the relevant properties of our complementary numerical simulations. 
 For further details on each code/simulation, we refer the interested reader to the respective core numerical papers (\citealt{Gaspari:2012a} and \citealt{2017MNRAS.464.4448W}).
 \subsection{Eulerian simulation with \flash} \label{sec::flash}
 The Eulerian simulation setup used in this work was first presented in \citeauthor{} \citeauthor{Gaspari:2012a} (\citeyear{Gaspari:2012a} -- G12), with the goal of studying the long-term evolution of self-regulated kinetic AGN feedback. The simulations are carried out with the \texttt{Flash4} adaptive-mesh-refinement code \citep{2000ApJS..131..273F}. By using the three-dimensional Euler conservation equations of hydrodynamics (Sec.~2 in G12), they model a typical brightest cluster galaxy (BCG) within a cool-core cluster halo with virial mass $M_{\rm vir}\sim 10^{15}\ \Msun$ and temperature $T_{\rm vir}\sim 5.5\ {\rm keV} \simeq 6\times 10^7\ {\rm K}$ (akin to Abell 1795). 
 In addition to the Euler equations, the main included physics is the source terms related to the plasma radiative cooling and mechanical AGN feedback. The radiative cooling $\approx n^2\Lambda$ (where $n$ is the gas density and $\Lambda$ the cooling function; \citealt{Sutherland:1993}), which is integrated with an exact explicit solver, induces the loss of temperature/pressure in the gaseous atmosphere (from $T\sim5\times10^7\ {\rm K}$ down to $10^4\ {\rm K}$). In the unbalanced regime, this would rapidly induce a catastrophic, radial cooling flow towards the central AGN. However, the SMBH responds back by injecting momentum and kinetic energy via recurrent jets/ultrafast outflows in the nuclear region, hence generating a counterbalancing heating mechanism affecting up to $r\sim100$ kpc (the details of the self-regulated AGN feeding and feedback are described below in Sec.~\ref{sec::AGN}).

 The global domain covers a volume of $(1.3\ \Mpc)^3$ sampled with 10 levels of concentric static meshes (SMR), which are refined by a factor of $2\times$ between each other along each dimension (the evolution is advanced by using only the timestep at the finest level, for maximum accuracy). In this work, we focus on the central $(170\ \kpc)^3$, i.e., the cluster core that includes the BCG with the central AGN and its impact region. The maximum level has a resolution $\Delta x  \simeq 300 \ \pc$. Here, we focus on a $100 \ \Myr$ period of the $5\ \Gyr$ simulation of G12. However, we use a much finer time output with a temporal resolution of  $\Delta t \simeq 0.1\ \Myr$. As shown in Sec.\,\ref{sec::results_grid}, this period covers a dozen (strong and weak) AGN outburst events, which properly sample the recurrent evolution of self-regulated AGN feedback. We randomly selected such interval in the core of the long-term evolution. However, we inspected other 100 Myr sub-periods, finding analogous results as shown here, corroborating the self-similar nature of the feeding and feedback loop. 
 
 \subsubsection{AGN feeding and feedback} \label{sec::AGN}
 Given the central role of AGN feeding and feedback, we outline here the related modeling and main features. While we describe the essential numerical ingredients, we refer to G12 for an extensive analysis of the AGN feedback properties and imprints not tackled here (e.g. the evolution of the radial profiles and cross-sections/maps of the thermodynamic variables).
 
 Self-regulated AGN feedback is modelled via the well-known relativistic rest-mass energy rate equation: 
 \begin{equation}\label{Pjet}
 P_{\rm jet} = \varepsilon_{\rm m}\,\dot M_{\rm feed}\,c^2, 
 \end{equation}
 where the feeding rate $\dot M_{\rm feed}$ is computed directly from the simulated gas accreting within the `nuclear' region ($r<500\ {\rm pc}$). As shown in G12, the accreted gas mass results to be almost entirely composed of cold gas ($T < 5\times 10^5$ K). Such cold clouds arise from the top-down multiphase condensation triggered within the turbulent hot halo, also known as CCA rain (Sec.~\ref{sec::intro}). The mechanical efficiency $\varepsilon_{\rm m}\approx 5\times10^{-3}$ used here is the optimal macro efficiency found to substantially counterbalance cooling flows in massive halos such as galaxy clusters (see also \citealt{Gaspari:2017_uni}).
 
 The second key component of proper self-regulation is the AGN feedback response. The triggered bipolar jets are injected with a thin `nozzle' approach, i.e., via opposing internal boundaries in the middle of the domain (with a one cell cylindrical radius). The nozzle jet flux rates (mass, momentum, and kinetic energy) are based on Eq.~\ref{Pjet} and $\dot M_{\rm out} =  2\,P_{\rm jet}/v^2_{\rm jet}$. 
 The initial, nuclear jet velocity is fixed at $5\times10^4\ {\rm km\,s^{-1}}$, motivated by observations of ultrafast outflows (e.g.~\citealt{Tombesi:2013}) and entrained radio jets (e.g.~\citealt{Giovannini:2004}). As the outflow propagates outwards, the jet/outflow loads more ambient mass, decelerating by over an order of magnitude.
 In the simulation analyzed here (e.g. Fig.~\ref{fig::evo_grid}), the jet power oscillates between $P_{\rm jet}\sim10^{44}-10^{46}$ erg\,s$^{-1}$, hence with variable nuclear mass outflow rates between $\dot M_{\rm out} \sim 0.1-10\ {\rm \Msun\,yr^{-1}}$. In this simulation, the injected nuclear feedback always point toward the $z$-axis; however, previous AGN outburst events and turbulence injection create a highly chaotic inner environment, thus leading to a natural wobbling of the jets typically within a few kpc radii and half opening angles of $\sim$25$^{\circ}$. 
 In G12, we tested different values of the AGN feedback parameters (e.g., injection zones, velocities, inclination angles), finding that varying the details of how AGN jets are injected lead to a similar global evolution after a few self-regulation cycles, i.e. with macro outflows generating large X-ray cavities and shocks (up to 50\,-\,100 kpc) that counterbalance the impending cooling flow.
 The key shaping factor is the AGN jet power, which will be also the dominant driver of enstrophy, as shown in Sec.~\ref{sec::results_grid}-\ref{sec::results_tracers}.
 We note that, in this study, we are interested in the ensemble properties of the AGN feedback over several Myr, rather than matching observations of instantaneous X-ray bubbles or shock morphologies.
 
 In the past decade, among all the tested models, the CCA rain regulating mechanical feedback has proven to be that more robust and best-consistent with key observables. Among the most notable properties worth to highlight are the following: 
 (i) quenching the recurrent cooling flow by over 2 dex in a gentle manner, i.e. without a central entropy inversion, as shown by {\it Chandra} and {\it XMM-Newton} data (G12);
 (ii) quenching the X-ray spectra more vigorously toward the soft X-ray band (\citealt{Gaspari:2015_xspec});
 (iii) reproducing X-ray imaging disturbances, such as large-scale X-ray cavities/bubbles with cool rims and entrained metals (\citealt{Gaspari:2012b});
 (iv) reproducing the tightness and slope of the cluster/group scaling relations, such as the $L_{\rm x} - T_{\rm x}$ (\citealt{Gaspari:2014_scalings}); and
 (v) stimulating extended warm/cold filaments emitting in H$\alpha$, which then condense into compact molecular clouds raining onto the AGN (\citealt{Gaspari:2017_cca,Gaspari:2018}). 
 Here, we explore in more depth the kinematical aspects of this AGN feeding/feedback model.

 \subsection{Lagrangian simulations with \CRaTer}\label{ssec::crater}
  Unlike Eulerian codes, which are based on a control volume discretization, Lagrangian codes discretize the fluid into mass particles. Lagrangian particles can be used as `passive' tracers of the hydrodynamical flow.
  The combination of an Eulerian baseline and a Lagrangian tracer infrastructure allows us to exploit the accuracy of the former and versatility/speed of the latter. 
  Specifically for our study, this enables to precisely control each set of mass particles, where and at which time they are injected, while being able to repeat such procedure for multiple experiments and generations of particles. To do so, we use \CRaTer, a novel Lagrangian tracer code presented in \citet{2016Galax...4...71W,2017MNRAS.464.4448W}, to which we refer for comprehensive numerical details. \\
 \indent
 The core idea of \CRaTer \ is as follows: in post-processing, $N_p$ mass particles are injected on top of the output of the Eulerian simulation. The particles read out the underlying grid data by interpolating the associated values to the tracer position on the grid. Next, the particles are passively advected forward in time to the next snapshot (with sub-cycling in between), where they repeat the interpolation of the grid data. This loop is repeated for the desired period of the Eulerian simulation. At runtime, tracers are injected/removed following the mass inflows/outflows of the Eulerian simulation.  \CRaTer\ was originally developed to model cosmic-ray particles and, later on, turbulence in cosmological simulations of galaxy clusters \citep[e.g][]{2017MNRAS.464.4448W,2017MNRAS.471.3212W,2020MNRAS.495L.112W}. Hence, we made a few modifications to \CRaTer\ that are highlighted hin the following. \\
 \indent
 Due to the large density/mass range in the \flash\ simulation, we do not assign a fixed mass to the tracers. Each particle has its own mass depending on its location of initialization within the simulation box. In this work, the average particle mass is $\approx 4\times10^4\ \Msun$.
 In addition, to mimic the continuous mass injection from the AGN, we inject 10 tracer particles in each of the $2^3$ nuclear cells. Each of the injected particles obtains mass of $1/80$ of the mass injected by the AGN outflow. 
 In the \flash\ simulation, mass accretion onto the SMBH is modelled by removing the amount of nuclear accreted mass, $M_{\mathrm{feed}}$ (Sec.~\ref{sec::flash}). 
 We mimic the same behaviour with the tracer particles by counting the number of tracers, $N_{\mathrm{t}}$, which cross the nuclear region and whose velocity vector is pointing towards the SMBH. We estimate the accreted mass per tracer, $M_{\mathrm{feed}} / N_{\mathrm{t}}$, and remove this mass from each tagged particle. If the mass of a tracer becomes $\le 0$, it is removed from the simulation. Particles that move outside of the simulation box are removed from the computation, too. \\
 \indent
 At the start of each \CRaTer\ simulation, we inject a total of $N_p \sim 10^7$ tracer particles inside a sphere with radius of $\approx 40 \ \kpc$ that is centered around the SMBH. To obtain an initial higher mass resolution in the vicinity of the SMBH, we divide the sphere into three shells and we inject a larger number of particles per grid cell in the inner shells. More specifically in the innermost sphere, $r \le 10 \ \kpc$, we inject four particles per cell each carrying a fourth of the cell mass. In the intermediate shell, $10 \ \kpc \le r \le 20 \ \kpc$, we inject two particles per cell each having half of the cell mass. In the outer shell of the sphere, $20 \ \kpc \le r \le 40 \ \kpc$, we inject one particle per cell each carrying the mass of their host cell. 
 These particles are advected forward in time for $100 \ \Myr$. Following the injection and removal of particles, the number of tracers increases by a factor of $\approx 3$ to a total number of $\approx 3 \times 10^7$ at the end of the simulation. 
 At every timestep, we save the interpolated grid quantities, e.g. density or temperature, as well as the enstrophy and the different dynamical terms recorded by each tracer. 
 Saving the interpolated data at every timestep allows us to group individual particles based on their location at the end of the simulation, and thus to backtrace their physical evolution in time (see Sec. \ref{ssec::tracer_families}). \\
 \indent
 For the interpolation of particle velocities \citep[e.g.][]{2010A&A...513A..32V} and other grid quantities, we use a {nearest-grid-cell} interpolation. Albeit being slightly more diffusive than a {cloud-in-cell} interpolation, this method assures a proper treatment of the interpolation at the interface between the different levels of the nested SMR grids. At the interfaces, we compute the enstrophy and the different dynamical terms using the data of the coarser grid. \\
 \indent
 Time-wise, since the timestep between the \flash\ snapshots is larger than the CFL condition, we use sub-cycling to advect the tracer particles between subsequent \flash\ snapshots. This approach has been extensively tested and applied in previous works \citep[][]{myphdthesis,2017MNRAS.464.4448W,2017MNRAS.471.3212W}. Here, we further tested different parent integration timesteps and snapshots (e.g. $5\times$ higher/lower than the default $0.1\ \rm Myr$), finding convergent results as those shown in Sec. \ref{sec::results_tracers}. 
 %
%
 \section{Eulerian Analysis}\label{sec::results_grid}
 \begin{figure}
     \subfigure{\hspace{-0.48cm}
     \includegraphics[width = 0.5\textwidth]{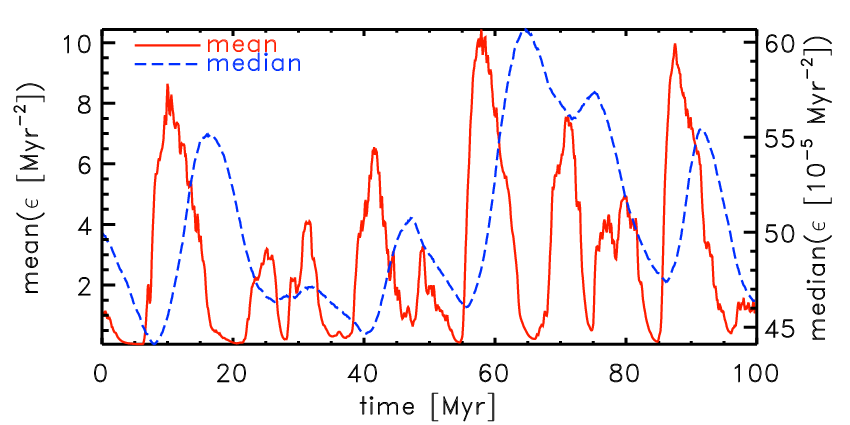}}
     \subfigure{\hspace{-0.48cm}
     \includegraphics[width = 0.5\textwidth]{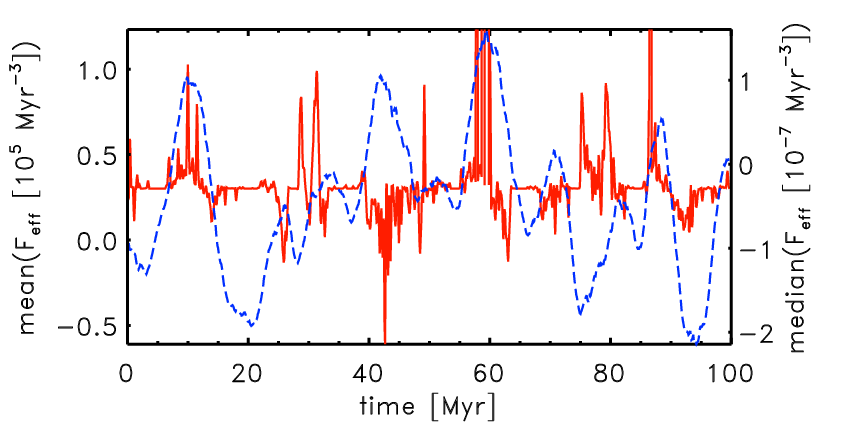}}
     \subfigure{\hspace{-0.48cm}
     \includegraphics[width = 0.5\textwidth]{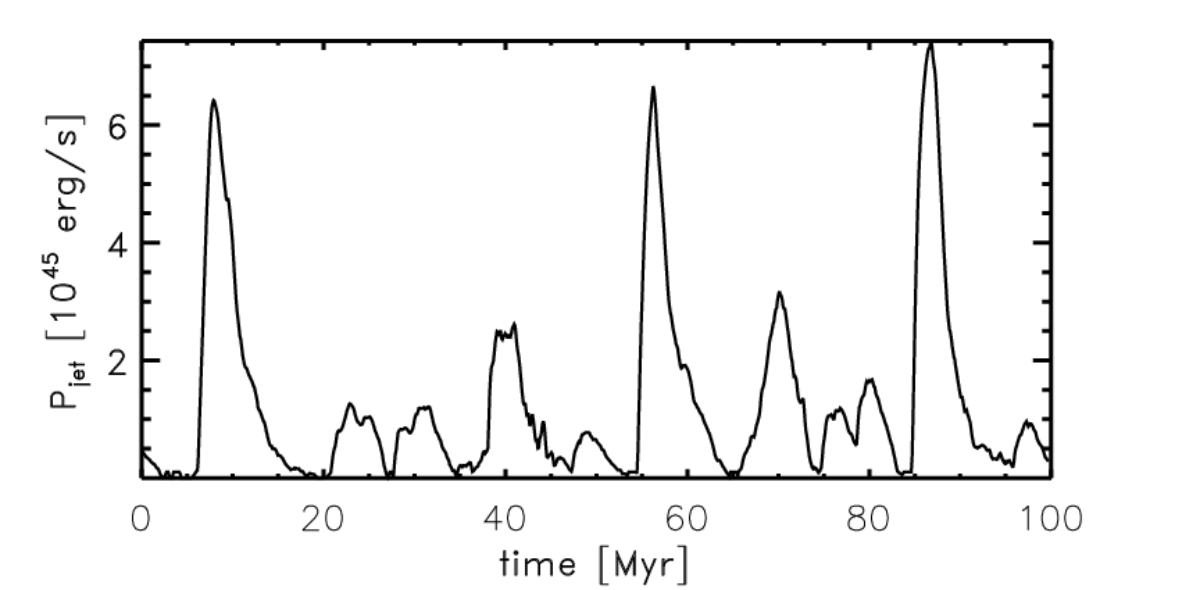}}
     \caption{Eulerian analysis: evolution of the enstrophy (top), the effective source term (middle), and the instantaneous mechanical AGN feedback power (bottom). In the top two panels, the red solid lines is the mean, while the blue dashed line is the median computed across the grid (notice the respective, different left/right $y$-axis labeling). To make the evolution of the mean effective term more visible, we limited its range to values below $ \sim 10^5 \ \Myr^{-3}$ (the peak at $\sim 60 \ \Myr$ reaches values of a few $10^7 \ \Myr$).}
     \label{fig::evo_grid}
 \end{figure}
 \begin{figure}
     \hspace{-0.99cm}
     \includegraphics[width = 0.54\textwidth]{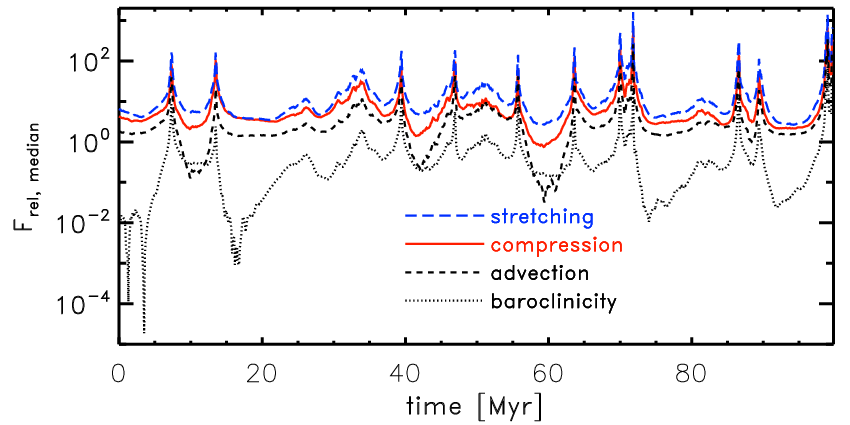}
     \caption{Eulerian analysis: median of the relative contributions of the different source terms (as absolute value) computed on the grid (see Eq. \ref{eq::frel}). Top: the red solid line is the compressive/rarefaction term, the blue long-dashed line is the stretching term and the black short-dashed line is the advection term. The black solid line in the bottom panel displays the relative contribution of the baroclinic term. We note that, as the effective term is the sum of both positive and negative values, each relative term $F_{\rm rel}$ is often $>1$; for similar reason, the related peaks can also differ from those in Fig.~\ref{fig::evo_grid}, \ref{fig::fcom_fstr_evo_grid}.
     }
     \label{fig::rel_evo_grid}
 \end{figure}
  \begin{figure}
     \subfigure{\hspace{-0.48cm}
     \includegraphics[width = 0.53\textwidth]{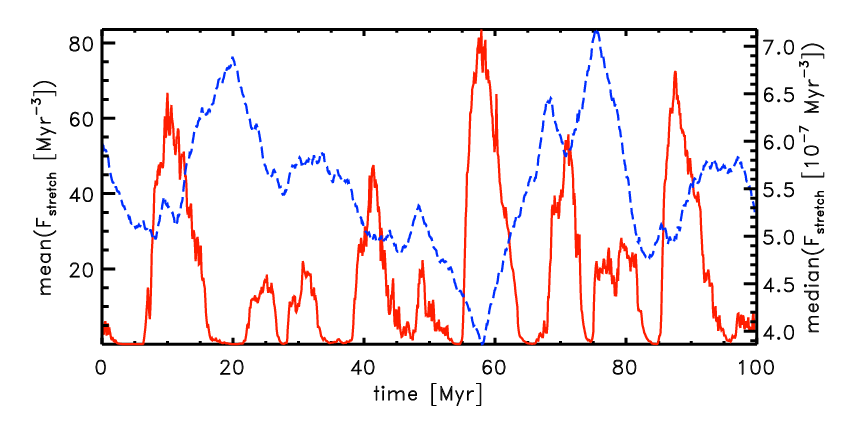}}
     \subfigure{\hspace{-0.48cm}     
     \includegraphics[width = 0.53\textwidth]{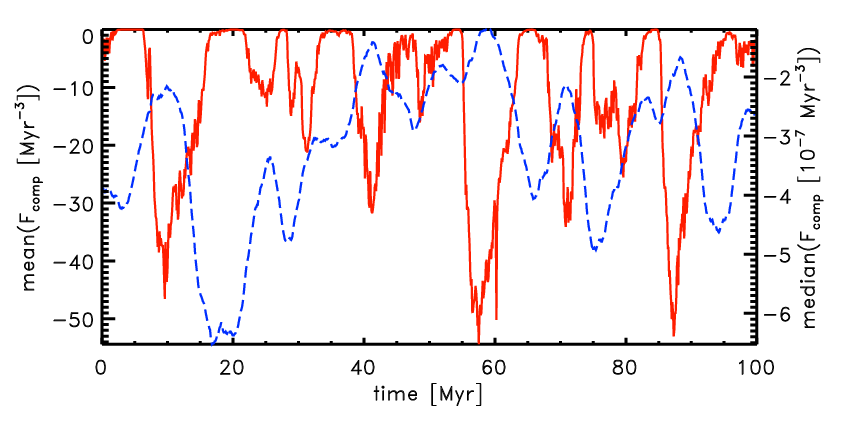}}
     \caption{Eulerian analysis: evolution of the mean (red solid) and median (blue dashed) of the stretching (top) and compressive/rarefaction (bottom) terms computed on the grid, including the dominant positive/negative range. (Notice the respective, different left/right $y$-axis labeling).
     }
     \label{fig::fcom_fstr_evo_grid}
 \end{figure} 

 As a first step, we analyse the evolution of enstrophy and the effective dynamical term in the Eulerian frame, see Fig. \ref{fig::evo_grid}. To indicate the AGN activity throughout the simulation, we further show the evolution of the injected mechanical AGN power $P_{\rm jet}$ in Fig. \ref{fig::evo_grid} (bottom panel). The overall thermodynamical framework is as follows (see also G12): the core ICM halo experiences quiescent phases of slight overcooling that drive the CCA rain towards the SMBH, quickly triggering self-regulated bipolar AGN outflows, which inflate a pair of cavities within an expanding elliptical cocoon (e.g.~Fig.~\ref{fig::f_time_seq550_650}, \ref{fig::f_time_seq850_950}). This triggers an intense phase of turbulent cascade along/within the cocoon, that is followed by a slower isotropization of the deposited AGN feedback energy over the $4\pi$ solid angle, with turbulent velocity dispersion oscillating between $\sigma_v \sim 80-700\ \rm km\,s^{-1}$.\\
 \indent
 Fig.~\ref{fig::evo_grid} shows that, with a short delay of $\lta$\,$2\ {\rm Myr}$, the mean enstrophy follows the active periods of the AGN/SMBH. They show the same number of local maximums, that can be divided into three strong and roughly six weak events. The relative amplitudes of the enstrophy maximums usually scale with the power of the AGN activity. Although sometimes, part of the kinetic energy is more quickly converted into thermal energy instead of vorticity (e.g., at $t\sim 90\ \Myr$).\\
 \indent
 The median of enstrophy (blue and dashed) is several orders of magnitude, $\sim 10^{-4}$, smaller than the mean (red and solid), indicating that the distribution of enstrophy on the grid is non-Gaussian and asymmetric. Indeed, the mean is dominated by relatively small regions that are strongly affected by the AGN activity, creating fat tails in the distribution. Conversely, the median is a good estimator of the properties of the bulk of the gas, here in terms of volume.
 A more careful comparison of the mean and median shows that the maximums of the median follow the peaks of the mean with a significant delay of $\sim 10 \ \Myr$, as the enstrophy/turbulence propagates more slowly over the bulk of the ICM atmosphere after the initial jet-inflated cocoon phase. \\
 \indent
 The mean and median of the effective dynamical term, i.e. the RHS of Eq. \ref{eq::enst_euler} (the net growth/decay rate), show analogous behaviour to that of the AGN power and enstrophy. The mean is several orders of magnitude larger than the median, by $\sim 10^{12}$, remarking that the mean is dominated by confined gas volumes that are majorly affected by the AGN activity. On the other hand, the  evolution of the two values follow roughly the same trend without any delay. \\
 \indent
 To investigate the importance of each dynamical term, we computed their relative contribution on the grid, Eq. \ref{eq::frel}. In Fig. \ref{fig::rel_evo_grid}, we plot the median of the relative contribution of each dynamical term. Both compression and stretching motions are the two dominant and comparable terms. Hence, they are the main driver of the enstrophy/turbulence evolution. Stretching tends to be slightly more persistent than compressions/rarefactions, preserving a low level of turbulence even during weak AGN feedback periods (more below).
 Advection of enstrophy is mostly subdominant and it becomes rarely comparable. Throughout the whole simulation, baroclinic motions are instead negligible; evidently, density and pressure gradients remain fairly well aligned during such a subsonic ICM turbulence. \\
 \indent
 As only compression and stretching motions determine the evolution of enstrophy, we plot their evolution in Fig. \ref{fig::fcom_fstr_evo_grid}, now considering signed quantities. As both quantities can be positive and negative, they can act either as a source or a sink of enstrophy. Both the mean and median show that stretching motions are always a source of enstrophy, i.e. amplifying it. On the other hand, the compressive motions are predominantly negative, i.e. the rarefaction mode is acting as a sink of enstrophy. Their peaks follow the evolution of AGN jet activity. Right after the jet-cocoon injection, the mean amplitudes of the stretching term are larger than that of the rarefactions. This leads to a local growth of enstrophy, which is gradually compensated by the rarefactions throughout the bulk of the volume. This shows another key dynamical self-regulation of AGN feedback turbulence -- highlighted further via the Lagrangian analysis in Sec.~\ref{sec::results_tracers}.\\
 \indent
 In summary, there is a strong correlation between the amplification of enstrophy and the AGN activity. Almost simultaneously with an AGN outburst, both stretching motions and rarefaction are amplified and they act as the main drivers of the enstrophy evolution. On the other hand, advection is rarely important and baroclinicity is always sub-dominant. Moreover, most of the amplification of enstrophy and the dynamical terms is fairly localized (more below). This supports the idea that the enstrophy evolution is closely connected to the evolution of impulsive AGN outbursts.
 \subsection{Zoom-in volumetric enstrophy evolution}
\begin{figure}
     \centering
     \includegraphics[width=0.49\textwidth]{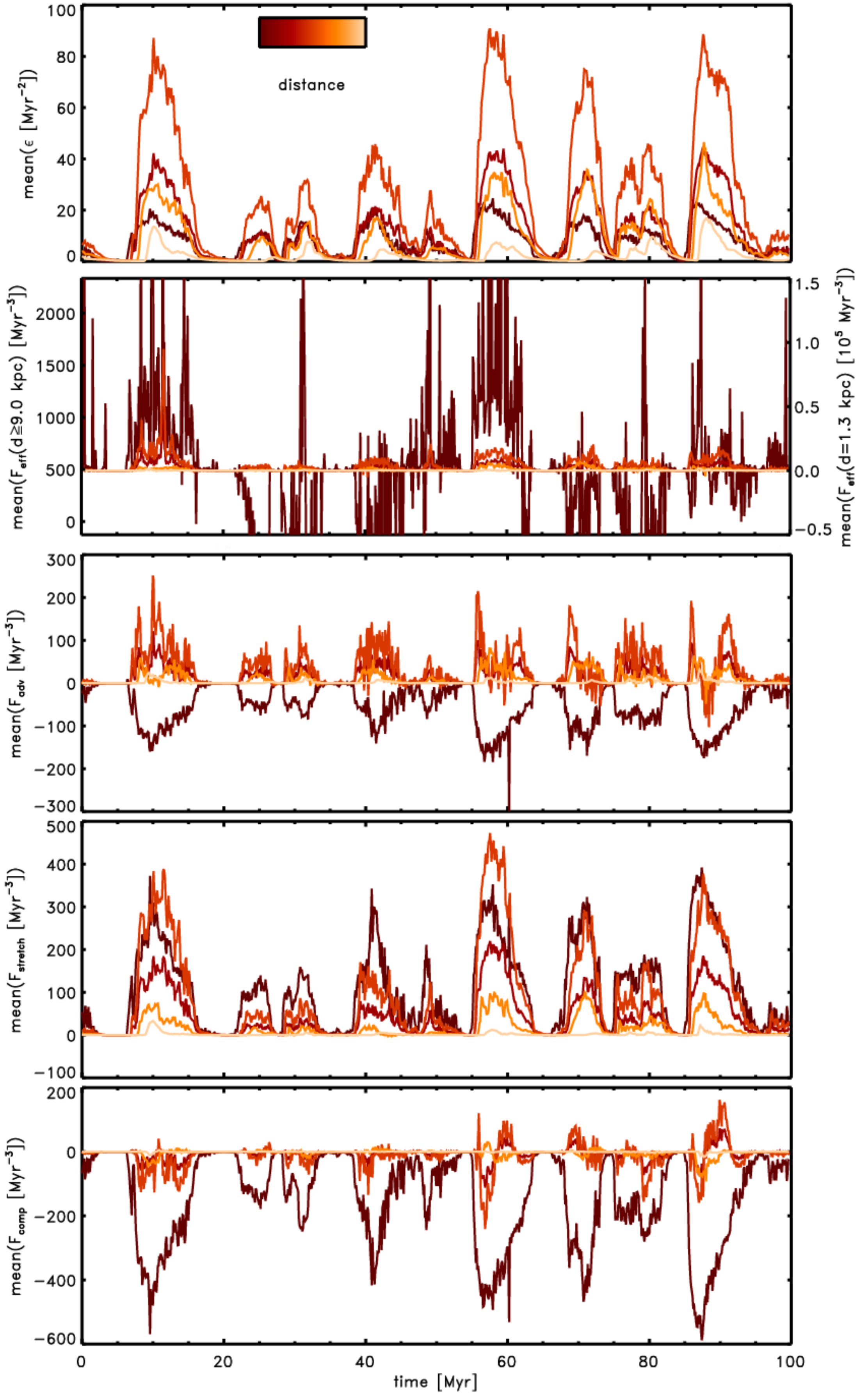}
     \caption{Eulerian analysis: evolution of the mean values of enstrophy (top row), effective dynamical term (second row), advection (third row), stretching (fourth row), and compression term (fifth row). The means are computed within pairs of short cylinders aligned with the bipolar jet path, with a cylindrical radius of $15 \ \kpc$ and increasing distance from the AGN ($\approx 1, 9, 18, 36, 72\ \kpc$; dark red to bright yellow colour). We point out that, due to the large spread in amplitudes of the effective term, the y-axis label on the right side of the plot gives the values for the innermost cylinder ($1.3 \ \kpc$ from the AGN).}
     \label{fig::eulerian_evo_shells}
 \end{figure}

\begin{figure*}
     \centering
     \includegraphics[width = 0.98\textheight, angle = 90]{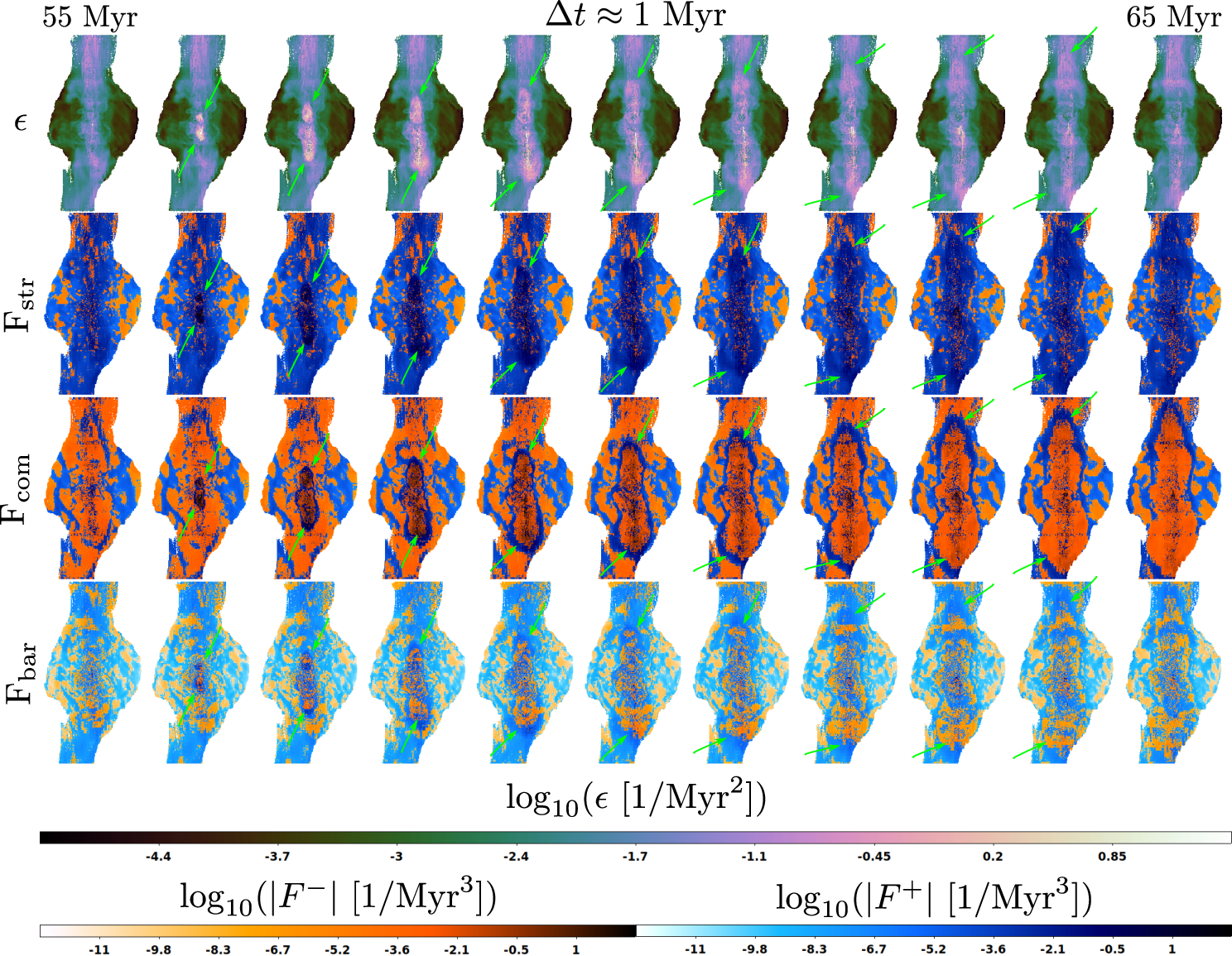} 
     \vspace{-0.5cm} 
     \caption{Lagrangian analysis: projected, mass-weighted enstrophy and dynamical sub-terms over the period $55-65 \ \Myr$, recording one of the strongest event of jet activity (Fig. \ref{fig::evo_grid}). The arrows highlight regions of maximum enstrophy.}
     \label{fig::f_time_seq550_650}
 \end{figure*}

 To further investigate the local spatial dependence of enstrophy and the dynamical terms, we compute their evolution inside five pairs of short cylinders aligned with the bipolar jet path. Each cylindrical radius is $r \simeq 15 \ \kpc$. The paired cylinders are located at the same radial distance on the opposite sides of the AGN. The first two pairs of cylinders are located in the most refined grid at radial distances (cylinder center to AGN) of $1.3 \ \kpc$ and $9.0 \ \kpc $, respectively, both with a cylindrical height of $2.5\ \kpc$. The other three pairs are located at radial distances (cylinder center to AGN) of $17.9 \ \kpc$, $35.9\ \kpc $, and $71.7 \ \kpc$, with cylindrical heights of $4.9 \ \kpc$, $9.9 \ \kpc$, and $19.8 \ \kpc$ (given the coarser grids), respectively. In Fig. \ref{fig::eulerian_evo_shells}, we plot the evolution of the mean values within each pair of cylinders. Here, we do not show the baroclinic term since it is dynamically unimportant. \\
 \indent
 We find that enstrophy (first panel) is always the strongest in the two pair that is located at an intermediate distances from the AGN, i.e. at $\sim 17.9 \ \kpc$. Hence, enstrophy is mostly amplified around the BCG region ($r\sim10-30\ \kpc$). Then it is transported further away from the AGN,  with only mild amplification, before becoming negligible at large $\sim 100\ \kpc$ radii. The effective dynamical term shows a similar global trend (second panel). Its magnitude in the two inner pairs is similar. However, in the innermost cylinder, the sign of its mean value is negative, which is due to the strong amount of rarefactions induced by the inner ultrafast outflow (fifth panel; remind that unlike in the above section, here we are analyzing very localized control volumes along the outburst path). 
 At tens of kpc from the AGN, stretching motions (fourth panel) overcome rarefaction and amplify enstrophy. Advection plays a minor role in evolving global enstrophy (third panel), as it quickly moves enstrophy from one place to another in fairly balanced way. Thus, it acts as a rapidly flipping source and sink term. 
 We note that Fig.~\ref{fig::eulerian_evo_shells} emphasizes the peaks and strongest enstrophy region (via the mean statistics) having rapid growth timescales; however, over the bulk of the volume, the effective rates are substantially lower (Fig. \ref{fig::evo_grid}), hence leading to larger growth timescales ($\gg$ Myr).
 \\
 \indent
 Overall, the above findings corroborate that the cylindrical jet is a very strong localized source of both stretching and rarefactions at short and medium distances from the AGN, $ <30 \ \kpc $. Enstrophy is amplified via stretching motions while it is transported away from the AGN in bipolar manner, until the rarefactions take over in the whole ICM volume (Fig.~\ref{fig::rel_evo_grid}). The latter leads to the significant decay of turbulence, as the jet power subsides between each major AGN feeding/feedback cycle (Fig.~\ref{fig::evo_grid}). 
\section{Lagrangian Analysis}\label{sec::results_tracers}
\indent

 Following Sec. \ref{ssec::crater}, we evolve up to $\sim$\,30 million tracer particles on top of the \flash \ data using \CRaTer. This allows us to study the fine details of the AGN chaotic weather. In the top row of Fig. \ref{fig::f_time_seq550_650}, we show the mass-weighted (Eq.~\ref{eq::mass_mean}) maps of enstrophy in the period of $55-65 \ \Myr$. This period covers the time before and after the strongest event of enstrophy amplification (see Fig. \ref{fig::evo_grid}). It also includes a major bipolar jet launching that is visible in the enstrophy map. \\
 \indent
 Both the advection of particles and the amplification of enstrophy are enhanced  along the jet axis, maximally near the AGN jet spearheads/hotspots (green arrows). The maps reveal that enstrophy is first generated in the center close to the SMBH and it is transported outwards along the jets. During the transport, enstrophy is amplified before it starts to diffuse. In the equatorial region, enstrophy is significantly lower and shows only minor variations. The spatial distribution of enstrophy is not symmetric with respect to neither the jet axes nor the equatorial plane. This reflects the anisotropic nature of the wobbling AGN jets, which are continuously precessing due to the long-term chaotic weather (cf.~G12). \\
 \indent
 To investigate the spatial distribution of the dynamical terms\footnote{We note that, in the Lagrangian frame, the compression term is $2\fc$, Eq. \ref{eq:Fcomp}, which is the quantity that we plot and analyze in this Section.}, we show their maps in Fig. \ref{fig::f_time_seq550_650} (second to fourth rows). In the blue regions of the maps, the individual dynamical terms act as sources and, in the red regions, they act as sinks. At each time step, the stretching term (second row) is positive throughout most of the volume and. Mostly in the equatorial region, it becomes negative in a few zones. However, these regions have low enstrophy. The dominance of vortex stretching over vortex squeezing is a common feature of subsonic hydrodynamical turbulence: in the full incompressible regime, the lengthening of the fluid/turbulent element implies thinning due to volume conservation.\\
 \indent
 As counterbalance, the compression term is negative (red) throughout most of the volume indicating motions of rarefaction. In most places, the sign of the compression and stretching term are the opposite, equilibrating each other, as suggested by the above Eulerian analysis. A remarkable feature in the maps of the compression term is the outwards expanding cocoon of compressions (the major AGN shock) that confines a large internal volume of strong rarefaction. Spatially, this envelope correlates with the tip of the region of amplified enstrophy (green arrows). The only other regions of positive compression lie in the equatorial region, where enstrophy is low, a sign of ringing sound waves (e.g., see the Perseus cluster; \citealt{Sanders:2007}). \\
 \indent
 Throughout the whole volume, baroclinicity is smaller than the other two source terms and its sign is not confined to specific regions, flickering rapidly in nearby zones. Nevertheless, the launched jet enhances the baroclinic term again slightly more along the jet axes. Evidently, the gentle AGN feedback is able to modify both the pressure and density gradient fairly coherently, despite the globally stratified ICM (in both $n$ and $T$). As analogy to Earth weather, this would compare to barotropic zones, i.e.~those found toward central latitudes, in contrast with polar regions suffering stronger cyclones. \\
 \indent
  We observe a similar behaviour around the other two events of maximum enstrophy and AGN power: namely in the periods $7-17 \ \Myr$ (Fig. \ref{fig::f_time_seq070_170}) and $85 - 95 \ \Myr$ (Fig. \ref{fig::f_time_seq850_950}).  
  The associated maps are shown in the appendix App.~\ref{app::more_maps}.
  While the global evolution is similar to the above description (a jet injection, followed by a cocoon and bipolar hot spots driving a large stretching zone, leading to rarefactions and expanding weak/transonic shocks), it is important to notice how variable and asymmetric the jet axis and cocoon are. While a single cycle is substantially anisotropic, the total AGN feedback energy is nearly isotropically re-distributed over tens of AGN events. This sustains an average mild background level of subsonic turbulence ($\rm Mach \sim 0.1-0.3$), which is often found indirectly via X-ray surface brightness fluctuations and optical H$\alpha$-emitting filament observations (e.g.~\citealt{Walker:2015, Hofmann:2016, Gaspari:2018, Simionescu:2019}). \\
  \begin{figure*}
     \centering
     \includegraphics[width = \textwidth]{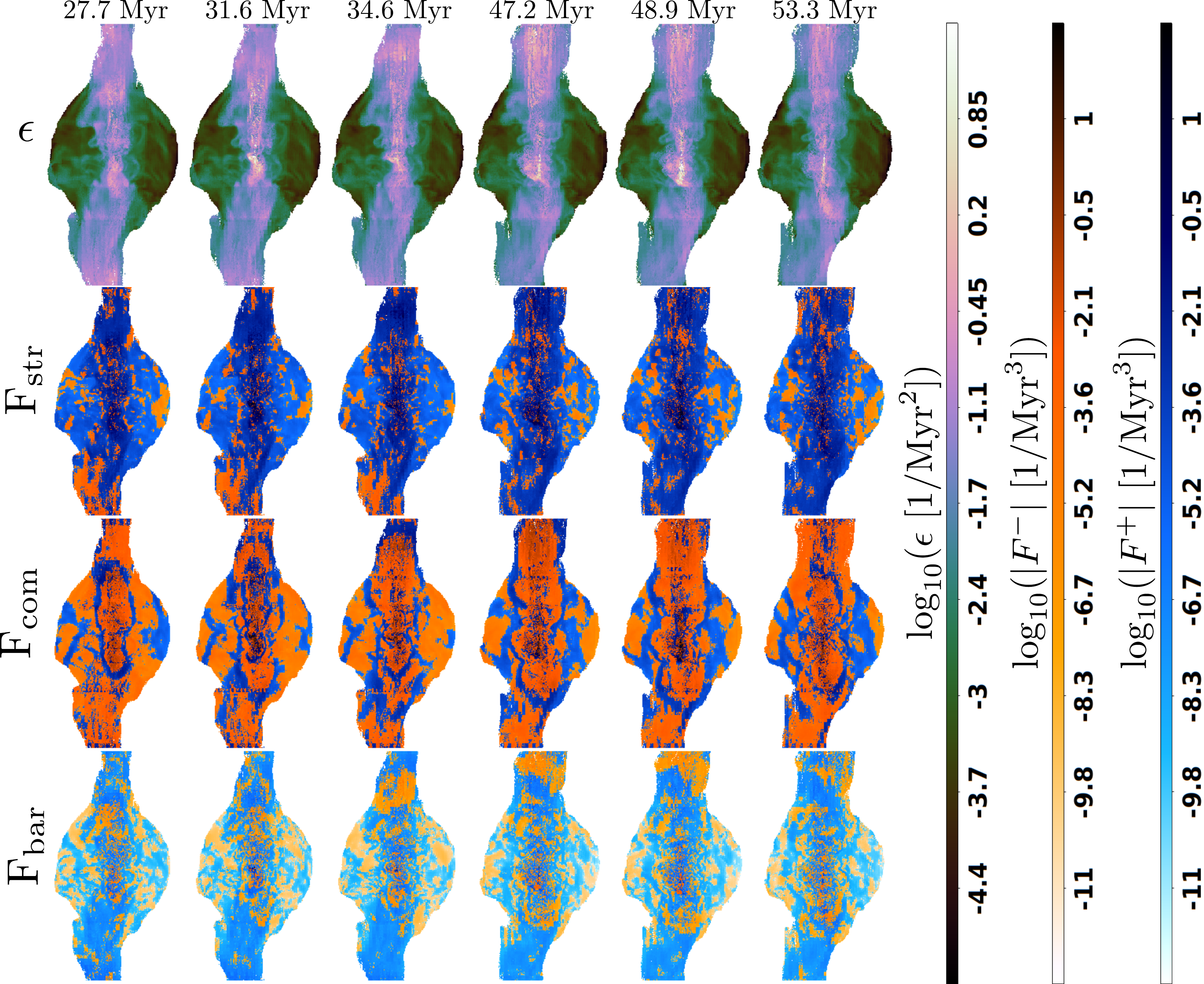} 
     \caption{Lagrangian analysis: projected enstrophy and dynamical terms recorded during two periods of low AGN activity (analogue of Fig.~\ref{fig::f_time_seq550_650}).   
     }
     \label{fig::f_time_seq_low}
 \end{figure*}
 \indent
 Further, during periods of lower AGN activity and enstrophy, the enstrophy and the dynamical terms show a behaviour analogous to the active phases. In Fig. \ref{fig::f_time_seq_low}, we show the same kind of maps as above for the periods $27.7 -34.6\ \Myr$ and $47.2 - 53.3 \ \Myr$, in which $P_{\rm jet}$ is $\sim5-6\times$ lower than that of the major outbursts. While previous feedback features are redistributed isotropically, there are still weaker AGN jet/bubble events which superpose further enstrophy along the jet cylinder.  Nevertheless, the relative importance of the dynamical terms is similar as in the majorly active phases of the AGN: within the expanding cocoon, stretching motions are mainly positive (blue) and act as a source. On the other hand, compression remains mostly negative (red) and, hence, serves as a sink. During more quiescent times, we observe a more disrupted morphology of the cocoon and bubble structures, making it more difficult to discern them from the background level of fluctuations/turbulence. The sign of baroclinicity changes again on very small scales. \\
 \indent
 Overall in line with the Eulerian analysis, the processes that determine the evolution of enstrophy are self-similar during active and quiescent times. Mainly the AGN jet power regulates the high/low turbulence states (which is, in turn, self-regulated by the CCA feeding).
 \subsection{Sixteen families of tracer particles}\label{ssec::tracer_families}
\begin{figure*}
    \includegraphics[width=1.02\textwidth]{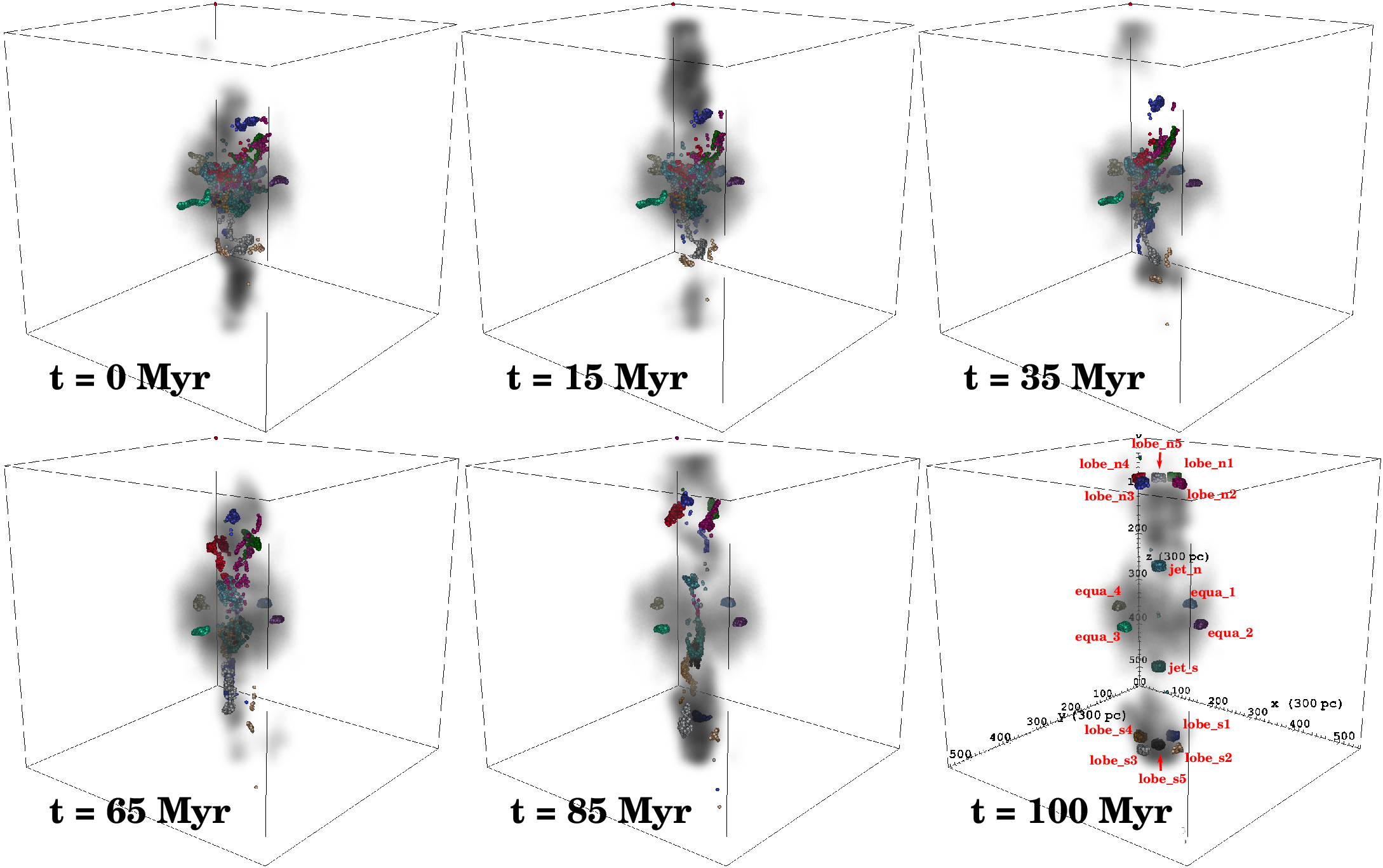}
    \caption{Lagrangian analysis: gas density (grey color; code units) and a sub-selection of tracer particles at six different timesteps of the simulation. 
    At $t = 100 \ \Myr$, the 16 families were selected according to their final position in small spheres, as labelled in red. The positions and properties at the previous timesteps are then backtraced (we remind that the full \CRaTer\ integration is still done forward in time). The units of the axis are in $\Delta x = 300 \ \pc$ and are only shown in the last panel.}
    \label{fig::visit}
\end{figure*}

\begin{figure*}
    \includegraphics[width = \textwidth]{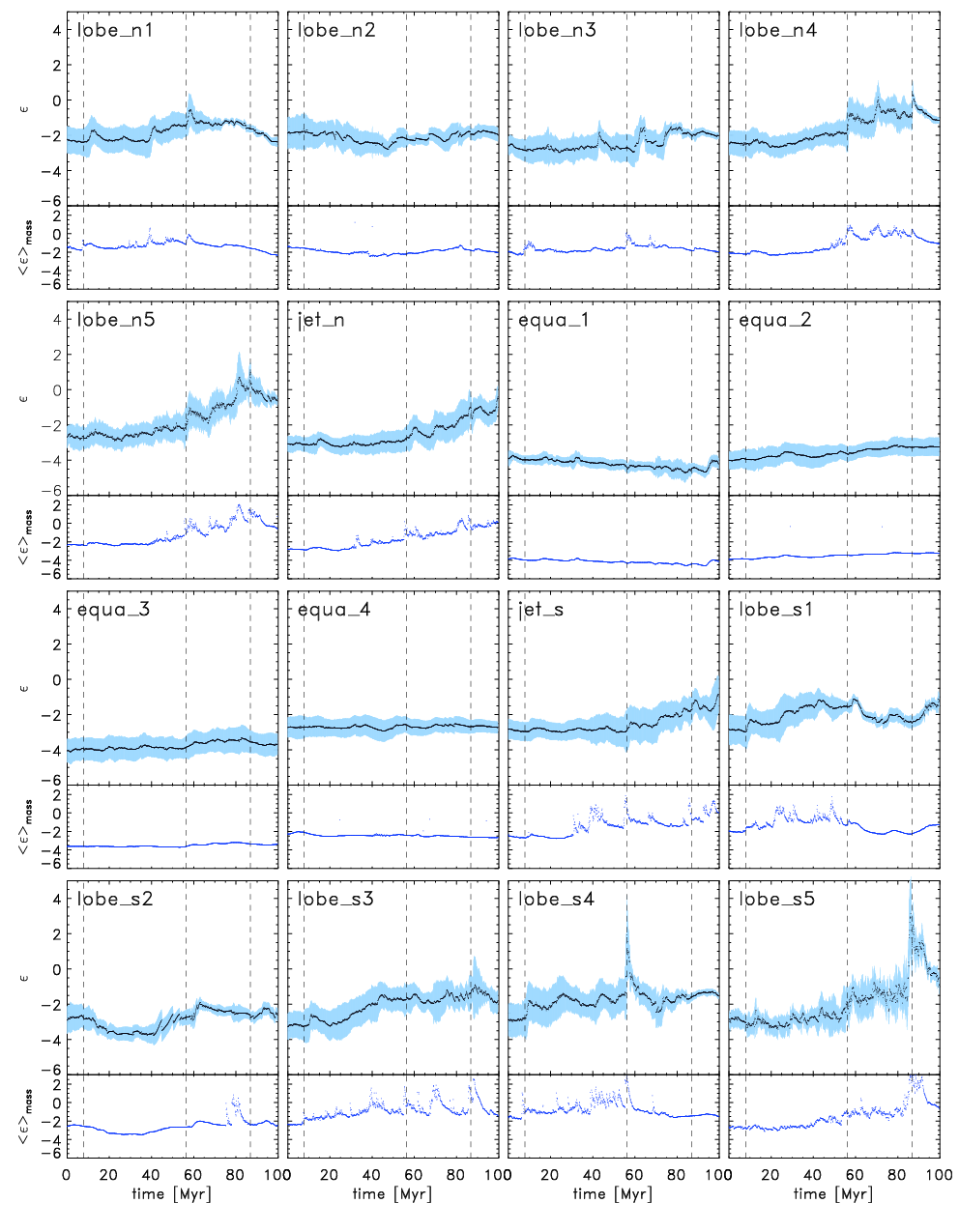}
    \caption{Lagrangian analysis: temporal evolution of enstrophy as recorded by the 16 different tracer families (Fig. \ref{fig::visit}). Each top/bottom panel show the median/mean of the recorded enstrophy, respectively. The blue bands mark $\pm 1$ standard deviation. The unit of the $y$-axis is $\log_{10}(\Myr^{-2})$. The three vertical dashed lines highlight the events of strongest enstrophy and AGN power (see Fig. \ref{fig::evo_grid}).}
   \label{fig::enstrophy_evo_4x2}
\end{figure*}

 At the end of the simulation, i.e. $t=100 \ \Myr$, we select all tracer particles that end up in 16 different spheres (with radius $3.4 \ \kpc$) located in various places around the AGN. The 16 regions are labelled in red in Fig. \ref{fig::visit} (bottom right panel). We refer to each selected sub-group as tracer `family'. Each family consists of  $N_p \approx 2\times10^2-2\times10^3$ particles. 
 In Fig. \ref{fig::visit}, we visualise their trajectories throughout the simulation, superposed to the gas density. 
 Families with ID name \texttt{lobe\_n1}--\texttt{lobe\_n5} are selected in the `northern' hemisphere, within the AGN bubble/lobe. Families \texttt{lobe\_s1}--\texttt{lobe\_s5} are located at the same positions but in the `southern' hemisphere. Finally, we select four families (named \texttt{equa\_1}--\texttt{equa\_4}) with a final location in the equatorial region (perpendicular to the jets), and two families (\texttt{jet\_n} and \texttt{jet\_s}) that are located inside the intermediate jets. \\
\indent
 We use the data recorded by the tracers to investigate the evolution of enstrophy and the associated physical processes recorded by each family (Eq. \ref{eq::enst_lagrange}). We employ a variety of proxies that are defined as follows.\\
 \indent
  As first step, we compute the mass-weighted mean of the enstrophy and the dynamical terms. For reference, the mass-weighted mean of a quantity $x$ is
 \begin{align}
     \langle x \rangle_{\mathrm{mass}} = {\sum\limits_{i = 1}^{N_p} m_i x_i}/{\sum\limits_{i = 1}^{N_p} m_i }, \label{eq::mass_mean}
 \end{align}
 where the sums are taken across the number of tracers per family, $N_p$, and $m_i$ is the mass of the $i$-th tracer particle. \\
 \indent
 As seen in Sec. \ref{sec::results_grid}, the values of enstrophy and the dynamical terms span several orders of magnitude. Hence, a few localized tracers can bias the mean to larger values and, thus, the mean and the median might differ significantly, signaling non-Gaussian distribution deviations (e.g. fat tails). In this case, the mean mirrors the dominant or strongest physical process that a few particles are exposed to. In the following, we will refer to these particles as outliers. On the other hand, the median provides information about the values recorded by the bulk of tracers, i.e. those filling the majority of the ICM atmosphere. To understand the physical processes, which the bulk of particles are exposed to, we compute the mass-weighted distribution of enstrophy and the three dynamical terms for each family in logarithmic space at every timestep of the simulation. Since the dynamical terms, Eq. \ref{eq:Fcomp}-\ref{eq:Fbaro}, can act as both sink and source terms, we compute the mass-weighted distributions for the positive and negative values separately. Hence, we define the medians of the positive and negative part the distributions of the dynamical term $F_i$ as:
 \begin{align}
     F_i^+ &= \mathrm{median} \left[ D \left( \log_{10} F_i \right) \right], \ \ \ \mathrm{where} \ F_i > 0  \label{eq::median_plus}; \\
     F_i^- &= \mathrm{median} \left[ D \left( \log_{10} |F_i| \right)\right], \ \ \mathrm{where} \ F_i < 0. 
     \label{eq::median_minus}
 \end{align}
 Here, $D(x)$ denotes the the mass-weighted distribution of the quantity $x$. By definition, enstrophy is always positive, so we compute the corresponding medians using Eq. \ref{eq::median_plus}. Analogous to Eq. \ref{eq::median_plus} and \ref{eq::median_minus}, we compute the standard deviation for each distribution. In Fig. \ref{fig::enstrophy_evo_4x2}, \ref{fig::fstr_evo_4x2}, \ref{fig::fcom_evo_4x2}, \ref{fig::fbar_evo_4x2}, we plot the evolution of the median plus/minus the standard deviation of each distribution. In each plot, the blue/orange bands display the median of the positive/negative distribution (Eq. \ref{eq::median_plus}/\ref{eq::median_minus}), respectively. 
 In the bottom sub-panel of each plot, we provide the absolute value of the mass-weighted mean; again, a blue/red data point indicates a positive/negative value and, hence, the dynamical term acts as a source/sink. Finally, we show the relative contribution of each source term over all families in Fig. \ref{fig::frel_evolution_all}. 
 \subsubsection*{Evolution of enstrophy}
 The evolution of enstrophy varies among the families (Fig. \ref{fig::enstrophy_evo_4x2}). Most families inside the lobes show an increase/decrease of enstrophy as the major cavities are inflated/fade away. In particular, such decay is evident towards the end of the simulation, i.e. within the last $10-20 \ \Myr$. In particular, they start to trace the decay of enstrophy developing at large radii from the sustaining power of the AGN (compare with Fig. \ref{fig::eulerian_evo_shells}). Conversely, the two families inside the jets are located at positions where enstrophy is expected to be at its maximum and, indeed, they measure an increase in enstrophy during the same period, in correlation with the AGN jet power. Whereas, tracers living in the equatorial region do not show strong variations of enstrophy and both the associated mean and median values remain fairly steady.  \\
\indent
 To first order, the outliers and the bulk of each family show similar trends of enstrophy evolution, which increase and decrease simultaneously and mostly differ in their magnitudes. More specifically, inside the lobes and jets, the outliers record values that are all above $\approx 10^{-3} \ \Myr^{-2}$ and, only in \texttt{lobe\_s2}, they show lower enstrophy. Temporarily, these values even spike above $\approx 10^{2} \ \Myr^{-2}$. In a few remarkable instances, \texttt{lobe\_s4} and \texttt{lobe\_s5}, the enstrophy increases to a few $10^{3} \ \Myr^{-2}$, due to the stronger instantaneous jet hotspot. On the other hand, the bulk of particles (medians) measure significantly smaller enstrophy, mostly below $1 \ \Myr^{-2}$. 
 Only \texttt{lobe\_n4}, \texttt{lobe\_n5}, \texttt{lobe\_s4} and \texttt{lobe\_s5} show values that are above $1 \ \Myr^{-2}$. Hence, the enstrophy recorded by the bulk of particles is $10^2-10^3$ below the measurements of the outliers. This corroborates the picture of a stable background turbulence, boosted recurrently in bipolar cones (up to several $100\ \rm km\,s^{-1}$).
 Noticeably, the median recorded by the family \texttt{jet\_s} is a few $10^4$ smaller than the mean, signaling the presence of major outliers (given the violent intermediate-scale jet impact).\\
 \indent
 The scatter of the distributions resides mostly within $\pm (0.2 - 1.2)$ dex, regardless of the AGN power. For a few families that reside in the lobes, it can however increase above $\pm 1.2$ dex; e.g., for lobe\_s4 and lobe\_s5 the scatter reaches values of $\pm 2.4$ and $\pm 2.2$ dex, respectively. This corresponds to linear variations of factors of $100$. Conversely, at large radii, as enstrophy decays, the scatter tends to thin down. \\
\indent
 The values recorded in the equatorial region tell a more quiescent story. Here, the mean values of enstrophy lie within the range of $10^{-5} - 10^{-2} \ \Myr^{-2}$. Furthermore, the outliers and the bulk of particles record similar values of enstrophy, deviating just by factor of a few. Such milder differences are also reflected in the scatter of the distributions that is significantly smaller in the equatorial region. Here, the particles never record a standard deviation above $\sim 0.8$ dex, i.e. corresponding to variations in enstrophy below a factor of $6\times$. \\ 
\indent
 These findings support and expand the Eulerian results found in Sec. \ref{sec::results_grid}: enstrophy strongly propagates along the jet path and is amplified for a significant amount of time before it decays again, mimicking the self-regulated AGN feeding and feedback cycle. The particles in the lobes are already experiencing the ensuing decay of enstrophy. Whereas, the particles in the jets are located where the peak of enstrophy is expected and measured, in particular towards the end of the simulation with two close AGN outbursts. As the scatter of enstrophy is larger along the jet axis, the bulk of particles in the jets and lobes do not measure the same level of variation of enstrophy like the few outliers that dominate the mean. 
\subsubsection*{Evolution of the dynamical terms}
 \begin{figure*}\centering
  \includegraphics[width = \textwidth]{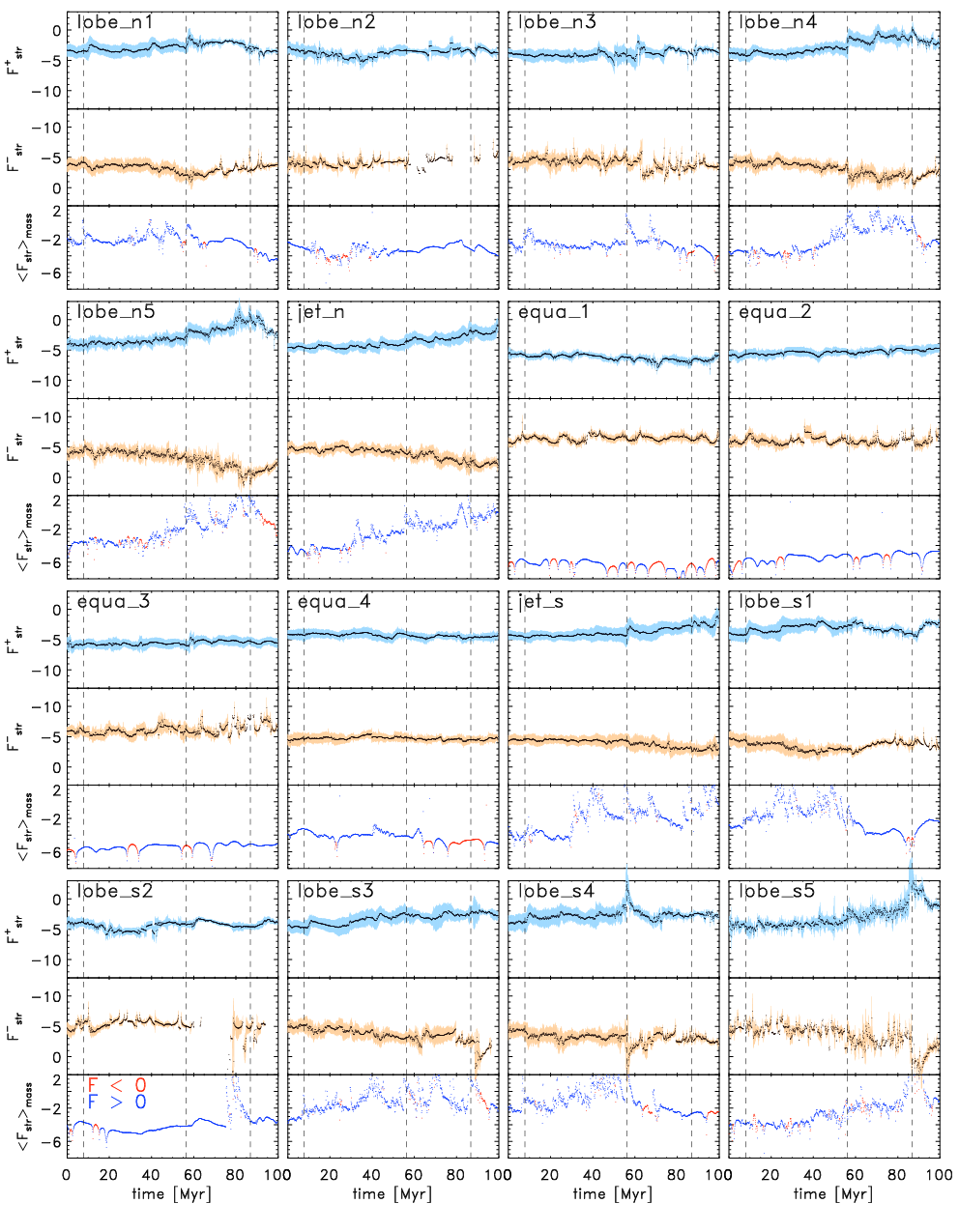}
  \vspace{-0.8cm}
  \caption{Lagrangian analysis: stretching term. Top and middle sub-panels: evolution of the median $\pm 1$ standard deviation of the positive (blue) and negative (orange) component of the stretching term measured by the 16 different tracer families (see ID labels in the top-left corner). Bottom sub-panel: mass-weighted mean stretching term measured by the tracers; a blue dot means the term is positive, while a red dot indicates a negative term. The unit of the $y$-axis is $\log_{10}(\Myr^{-3})$. The three vertical dashed lines mark the events of strongest enstrophy and correlated AGN power (Fig. \ref{fig::evo_grid}).}
  \label{fig::fstr_evo_4x2}
\end{figure*}
 \begin{figure*}
    \includegraphics[width = \textwidth]{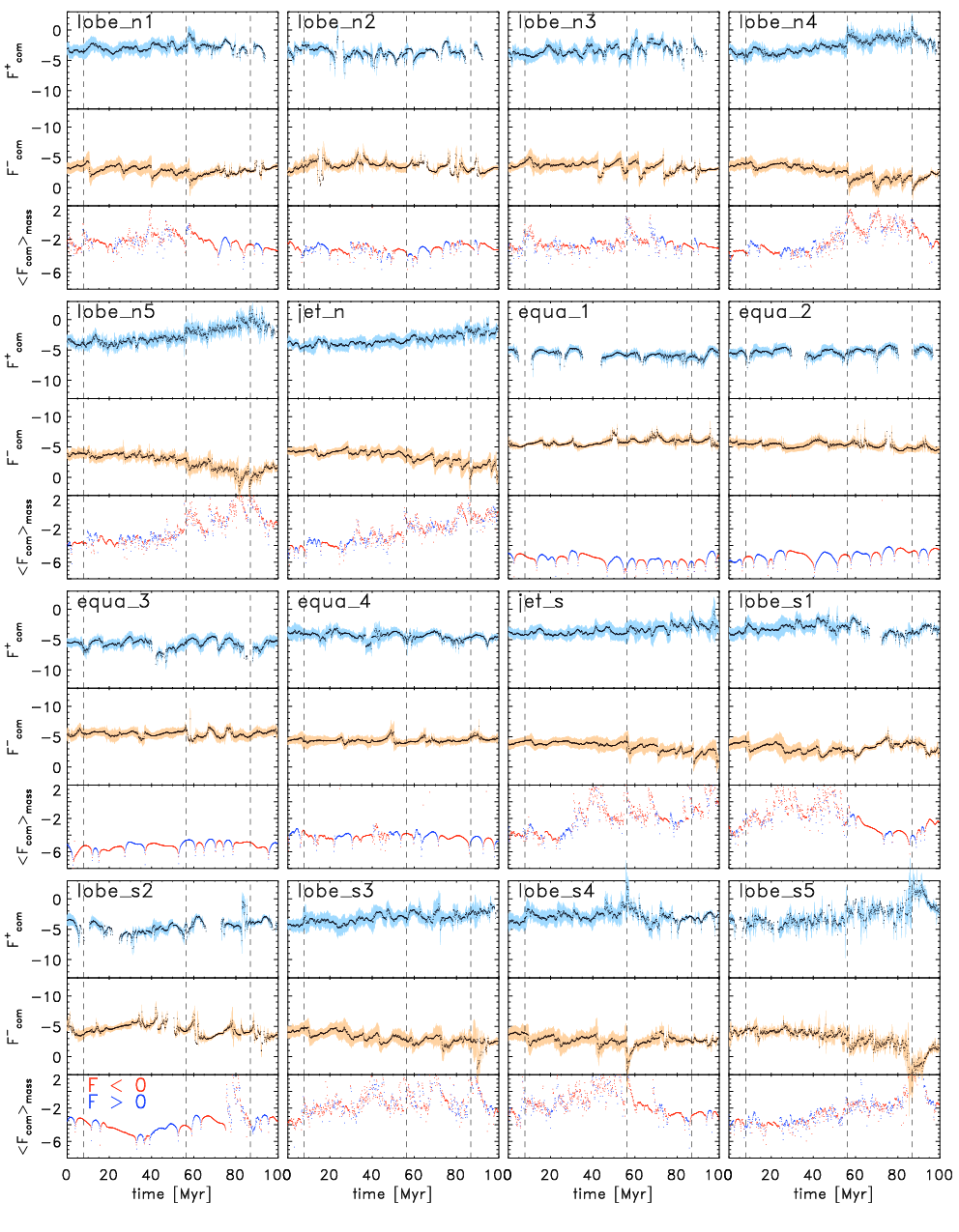}
    \caption{Lagrangian analysis: compression term. Analogue of Fig. \ref{fig::fstr_evo_4x2} for the compressive term measured via the 16 tracer families.}
    \label{fig::fcom_evo_4x2}
 \end{figure*}
 \begin{figure*}
    \includegraphics[width = \textwidth]{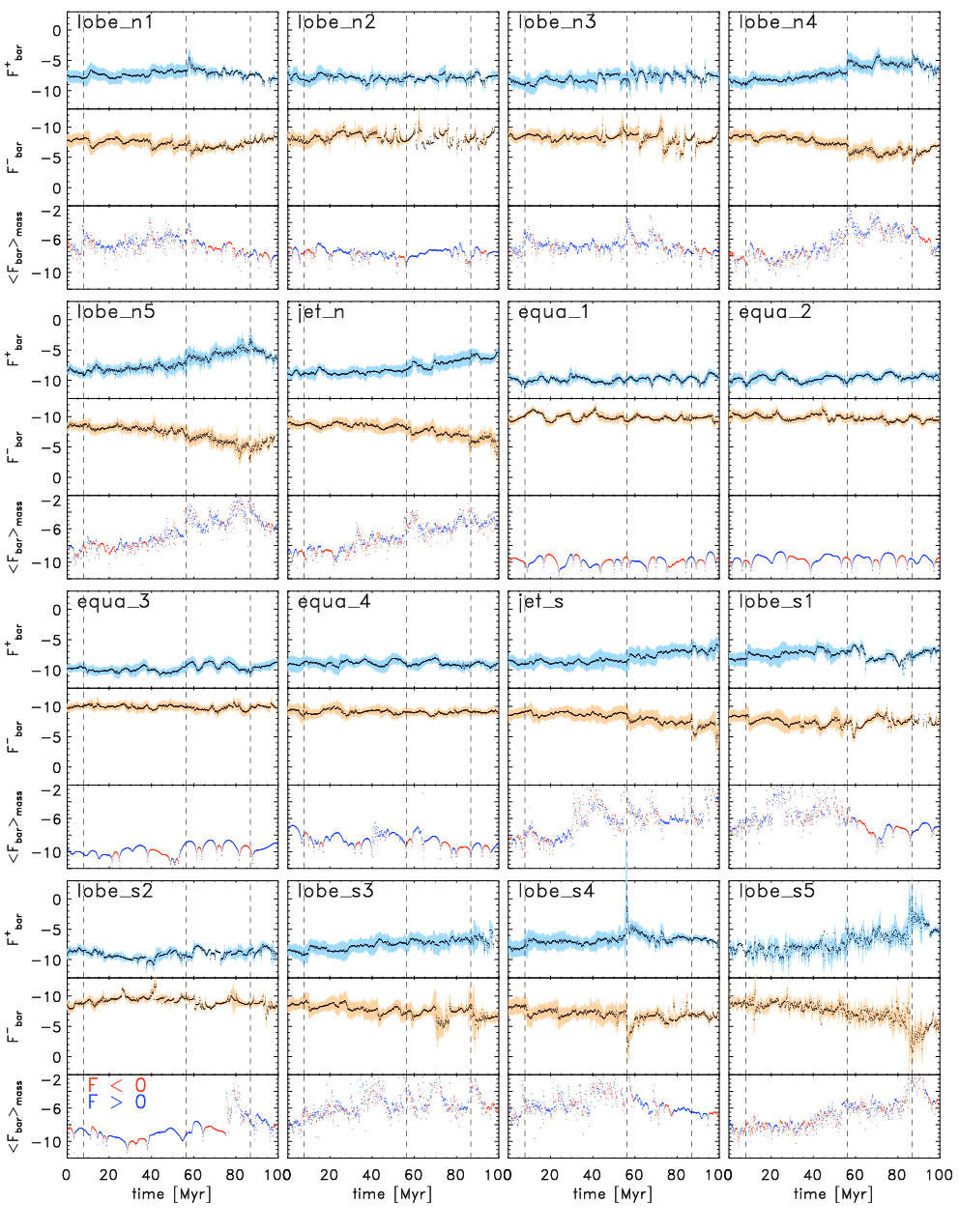}
    \caption{Lagrangian analysis: baroclinicity. Analogue of Fig. \ref{fig::fstr_evo_4x2} for the baroclinic term measured via the 16 tracer families.}
    \label{fig::fbar_evo_4x2}
 \end{figure*}
 \begin{figure}
     \centering
     \includegraphics[width=0.49\textwidth]{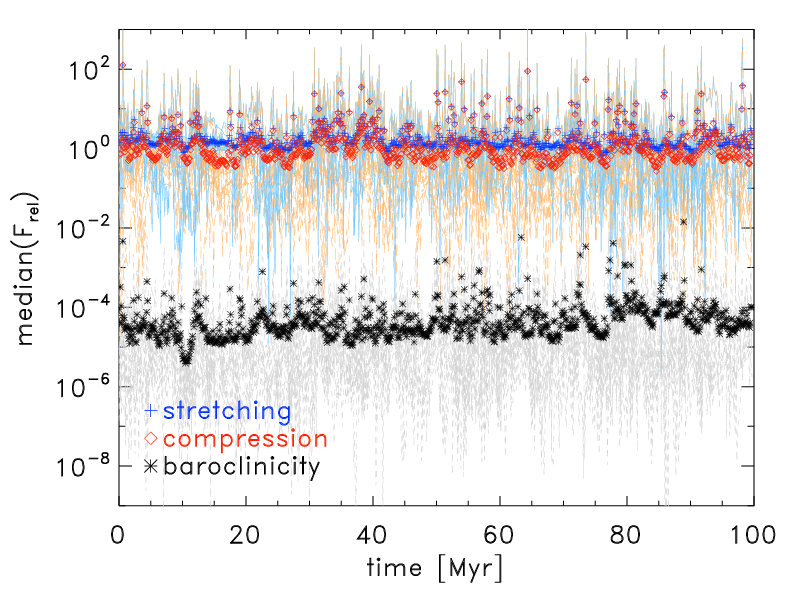}
     \caption{Lagrangian analysis: temporal evolution of the relative strength of the different source terms, Eq. \ref{eq::frel}. The light lines give the median measured by each tracer family. The superposed symbols give the average of each set of lines. The colours are similar as in Fig. \ref{fig::rel_evo_grid}: the blue lines/crosses are divergences (compressions or rarefactions), the red long-dashed lines/diamonds are stretching motions, and the black short-dashed lines/asterisks are tied to baroclinicity.}
     \label{fig::frel_evolution_all}
 \end{figure}
 As shown in Fig. \ref{fig::fstr_evo_4x2}-\ref{fig::fbar_evo_4x2}, the absolute value of each dynamical term (Eq. \ref{eq::enst_lagrange}) coherently increases/decreases, mirroring the enstrophy evolutionary trends. Again, this correlates with the main AGN power outbursts (vertical dashed lines), intervaled by weaker outflow/bubble events, which sustain a turbulence base level ($\sigma_v\sim 100\ \rm km\,s^{-1}$). Notice that, essentially at all times, the enstrophy evolution is a continuous competition between both positive and negative stretching/squeezing (Fig. \ref{fig::fstr_evo_4x2}), compression/rarefaction (Fig. \ref{fig::fcom_evo_4x2}), and baroclinic terms (Fig. \ref{fig::fbar_evo_4x2}), albeit one signed component can dominate over the other on average (hence, the importance of showing all separate signed components).\\
 \indent
 In the lobes and jets, stretching motions (Fig. \ref{fig::fstr_evo_4x2}) mostly amplify enstrophy (blue), while rarefactions (Fig. \ref{fig::fcom_evo_4x2}) reduce enstrophy (red). The particles residing close to the equatorial zone, measure stretching motions that are also recurrently negative (squeezing) and compressions that are also positive (shocks/sound waves), at variance with the jet/lobe families. This quantifies better the findings from the visual maps of Fig. \ref{fig::f_time_seq550_650}: in the equatorial region, the stretching and compression term flip their signs more frequently and, hence, both can act as sinks and sources in a short time span. This is due to the fact that such families experience gentler dynamics with a composition of weak transonic shocks, uplift/backflow, and turbulent mixing, compared to the violent concussions experienced along the jet path.\\
 \indent
 Indeed, we observe that the magnitudes of the stretching and rarefaction terms are significantly weaker in the equatorial region than in the lobes and jets, where the mean stretching is often as vigorous as $\sim 1 \ \Myr^{-3}$ and peaks at $10^2-10^3 \ \Myr^{-3}$. In the equatorial region, the mean value of the stretching term instead barely increases above $\sim 10^{-3} \ \Myr^{-3}$. Similar magnitudes and behaviour are seen for the rarefaction term. 
 In all families, the baroclinic term is highly chaotic, randomly changing sign and acting rapidly as a sink or source term, with no clear mean long-term up- or down-trend. Again, this shows that the dynamics of the baroclinic term do not depend on the environment and randomly vary on small spatial scales (Fig. \ref{fig::f_time_seq550_650}). Nevertheless, the low values of baroclinicity point out that the density gradient and pressure gradient are well aligned throughout the whole volume.  
 Hence, baroclinicity remains dynamically unimportant and does not determine the evolution of enstrophy in any of the families.
 This is corroborated by the fact that over the bulk of the volume most shocks are transonic, approaching sound waves, thus not having enough power to create complex misaligned gradients (similar concept applies to the gently quenched cooling flow). 
  \\
 \indent
 Independent of their functionality, the magnitude of the dynamical terms fluctuates more along the jet axis. Here, the bulk of particles (median) tends to show weaker dynamical terms than the associated outliers (mean). The related intrinsic scatter (blue/orange bands) can exceed the typical $\pm 1-2$ dex, indicating large variations of motions inside the impulsive jets and cocoon. In a few cases (see lobe\_s3, lobe\_s4, and lobe\_s5), the standard deviation is dramatically boosted up to $\pm 3$ dex, i.e., the strength of each particle source terms can differ by factors of up to $10^3$, even if spatially correlated. In the equatorial zone, the intrinsic scatter of the dynamical term fluctuates less prominently, remaining below $\pm 1.5$ dex, despite the rapid sign flickering. At variance with the cocoon region, in this zone the mean and median are often comparable, approaching smoother Gaussian distributions. The latter is a sign of less shocked dynamics.\\
 \indent
 For each family, we plot the median (light curves) of the relative contribution of the different source terms (Eq. \ref{eq::frel}) in Fig. \ref{fig::frel_evolution_all}. We further superpose the average behaviour of the families (symbols). As ensemble, the compressive and stretching motions are the key drivers of the chaotic AGN weather. Baroclinicity is always smaller and it merely contributes $ 0.01 \ \%$ to the effective dynamical term. Remarkably, the relative strengths of stretching and compression are similar throughout the whole duration. Again, this is a sign of an irreducible background level of turbulence, when averaged over large spatial (or temporal) scales, despite recurrent superposed boosts in the jet cones. In particular, as the turbulence cascades fill the whole volume, stretching motions are slightly more dominant and steady than rarefactions, while the bubbles/sound waves fade away.\\
 \indent
 Finally, the characteristic growth time (Eq. \ref{eq::turntime}) gives the timescale over which enstrophy is substantially modified by the effective source term. Averaging over all the families, we find an effective growth time of the order of 10-100 Myr (from the inner to outer radii), which is roughly comparable to the radial cooling time \citep[e.g.][]{Gaspari:2018}. Nonlinear thermal instability driven directly via turbulence is indeed one of the main mechanisms for the formation of CCA and precipitation in the ICM and intra-halo medium (e.g. \citealt{Gaspari:2013_cca, Voit:2017}). 
\subsection{Zoom-in tracer family analysis}
 The above findings point towards a global consistent picture of the evolution of enstrophy and its source terms. However, on smaller scales, there are differences and asymmetries between the measurements within each of the macro families (equatorial, lobes, jets) shown in Fig. \ref{fig::fstr_evo_4x2}-\ref{fig::fbar_evo_4x2}. Before concluding, it is worth discussing them further. We note that the retrieved chaotic behaviour (tied to turbulence stochasticity) remarks the difficulty of predicting the turbulent motions/weather injected by the AGN feedback in a given slice of space-time, unless in possession of the detailed boundary conditions in the very recent past, akin to Earth weather forecasts (\citealt{Bauer:2015}).
 \subsubsection*{Families within the equatorial region}
 Most families in the equatorial region show a fairly steady enstrophy evolution without any strong fluctuations. Yet, the enstrophy measured by \texttt{equa\_4} is significantly larger, with both the mean and median reaching $\approx 10^{-3} \ \Myr^{-3}$. This is $\sim 10\times$ larger than that measured by the other three families. \texttt{equa\_1} and \texttt{equa\_3} are located in the equatorial region at opposite sides of the AGN. In the first half of the simulation, their median enstrophy takes similar values. However, the enstrophy of \texttt{equa\_1} is significantly lower after $t \gta 50 \ \Myr$. Indeed, \texttt{equa\_1} shows the least amount of scatter and its recorded enstrophy gradually decreases over time, experiencing less turbulent mixing. These findings point out that, even in the equatorial region where there is no significant amount of enstrophy being generated, the spatial distribution of turbulence can be anisotropic, over short periods of time. \\
 \indent
 The positive and negative medians of the dynamical terms recorded by all four families are also fairly steady in time. Yet, the stretching term of \texttt{equa\_4} varies less than the ones of the other families. The compression and baroclinic terms, in particular, often act as both sources or sinks of enstrophy for these families, underlying the gentle but rapid variations occurring in the equatorial region. Indeed, the amplitudes of the dynamical terms do not show any major increase or decrease, with a few exceptions. For example, the stretching motions recorded by \texttt{equa\_2} increase by an one order of magnitude during the entire run. Another example is the baroclinic term recorded by \texttt{equa\_4}, which decreases by more than one order of magnitude during the simulation.
\subsubsection*{Families within the northern lobe} 
 There are substantial differences in the enstrophy evolution of the families that reside in the northern lobe. The enstrophy recorded by \texttt{lobe\_n1} and \texttt{lobe\_n3} is mainly affected by the first two strong outburst of the AGN at $8 \ \Myr$ and $56 \ \Myr$. Conversely, \texttt{lobe\_n4} and \texttt{lobe\_n5} show a significant increase of enstrophy after $\sim 60 \ \Myr$, i.e. right after the AGN outburst at $56 \ \Myr$ (see Fig. \ref{fig::evo_grid}). Subsequently, they are affected by the strong AGN outburst at $86.6 \ \Myr$. Strikingly, \texttt{lobe\_n2} is the only family that does not record any major amplification of enstrophy. This unveils the highly wobbling and precessing nature of the AGN jets/outflows, even along one direction.\\
 \indent
 The several dozen minor AGN events affect the evolution of enstrophy, too: \texttt{lobe\_n1} records amplification and variation of enstrophy at $\sim40$ and $50 \ \Myr$, which are associated with the concurrent outbursts. Even though, \texttt{lobe\_n3} records the same major AGN outbursts as \texttt{lobe\_n1}, it does not indicate any evolution of enstrophy associated with the intermediate minor AGN outbursts at $\sim 31$, $40$, and $48 \ \Myr$. \texttt{lobe\_n4} and \texttt{lobe\_n5} are the only families that show amplification of enstrophy directly connected to the minor events at $\sim 70$ and $80 \ \Myr$, closer tracing the hotspots of the driven AGN bubbles. These findings emerge in the the evolution of the different dynamical terms too: for all three terms the medians and scatter recorded by \texttt{lobe\_n4} and \texttt{lobe\_n5} fluctuate and increase after $\sim 60 \ \Myr$.  
 \texttt{lobe\_n2} and \texttt{lobe\_n3} measure fairly constant median values with small amplifications. Remarkably, the dynamical terms recorded by \texttt{lobe\_n1} dramatically decrease towards the end of the simulation. Hence, they are not affected by major AGN bubbles (unlike the opposite \texttt{lobe\_n4}). 
 \subsubsection*{Families within the southern lobe} 
 Similarly, the particles that reside in the southern lobe do not measure amplification and decay of enstrophy simultaneously: \texttt{lobe\_s1} and \texttt{lobe\_s4} record variations of enstrophy that is associated with the major and minor AGN outbursts between $10-60 \ \Myr$. In contrast, \texttt{lobe\_s2} and \texttt{lobe\_s5} measure an enhanced enstrophy evolution after $60 \ \Myr$, that is affected by several subsequent AGN events. Finally, \texttt{lobe\_s3} measures amplification of enstrophy during the entire run, mirroring each feedback event. Interestingly, the enstrophy uptrend in \texttt{lobe\_s5} is significantly boosted only at late times. This is connected to the fact that the particles of \texttt{lobe\_s5} were not aligned with the jet path at early times (we remind that the tracer selection is done at final time).\\
 \indent
 The dynamical terms display such differences as well: the stretching motions recorded by \texttt{lobe\_s1} and \texttt{lobe\_s4} are the strongest within the first $\sim 60 \ \Myr$ of the simulations, while, for the other three families, they grow towards the end. We observe \texttt{lobe\_s2} and \texttt{lobe\_s5} experiencing major stretching motions only once, while \texttt{lobe\_s3} undergoes four peaks of strong stretching (intersecting multiple AGN bubbles and cocoons). For all southern lobe families, the compression and baroclinic term evolve similarly to the stretching term, aside the different sign. 
\subsubsection*{Families within the lobe tips} 
 It is also worth to compare the evolution of \texttt{lobe\_n5} and \texttt{lobe\_s5} as they are located at the tip of the jet but on opposite sides of the AGN. Quantitatively, they measure the same evolution of enstrophy: at first the enstrophy is fairly constant ($\sim10^{-3}\ \Myr^{-2}$), before it starts to substantially increase after $40 - 50 \ \Myr$, as they align with the outflow path and drilled hotspots. However, \texttt{lobe\_s5} undergoes a much stronger episode of stretching and, thus, the maximum of its median enstrophy is much stronger than the one of \texttt{lobe\_n5}. This highlights the multiple asymmetries, not only between the cocoon and perpendicular region, but also between different hemispheres.
\subsubsection*{Families within the jets} 
The families \texttt{jet\_n} and \texttt{jet\_s} are located along the intermediate jet path on opposite sides of the AGN. Initially, the bulks of particles record similar values of enstrophy of $\sim 10^{-2}  \ \Myr^{-2}$. Towards the end of the simulation, the bulks of particles measure instead an enstrophy of $\sim 10^{-0.5} \ \Myr^{-2}$. The intermediate evolution of enstrophy is similar between the two families. Analogous behaviour applies to the median dynamical terms. The particle outliers reveal a slightly different picture: throughout most of the simulation, the outliers of \texttt{jet\_s} measure a larger enstrophy/turbulence. Between $\sim 30 - 70 \ \Myr$, the mean enstrophy of \texttt{jet\_s} is larger by a factor of a few, later increasing up to $100\times$. Indeed, the particles of \texttt{jet\_s} are often more aligned and thus more affected by the two AGN outflows at $\sim 31 \ \Myr$ and $\sim 40 \ \Myr$. On the other hand, \texttt{jet\_n} is more exposed to the outburst at $\sim 80 \ \Myr$. In \texttt{jet\_n}, all three dynamical terms show an overall constant increase with mild variations, whereas in \texttt{jet\_s} they show a large spike at $\sim 35 \ \Myr$ and smaller variations afterwards.\\
 \indent
Unlike the more violent and localized motions of \texttt{jet\_s}, the majority of the particles related to \texttt{jet\_n} are exposed to coherent flows. 
This proves that turbulent motions can be significantly different and intermittent, even deep within the strongest zones of the AGN feedback bipolar injection. 
 \section{Conclusion}\label{sec::conclusion}
 In this work, we studied the turbulent motions injected by AGN outflows/jets in the hot halo (within and around the BCG). We dissected the evolution of enstrophy, a robust proxy for solenoidal turbulence, and its dynamical terms (Eq. \ref{eq::enst_euler}-\ref{eq::enst_lagrange}) in a zoom-in SMR hydrodynamical simulation of AGN feedback, which has passed several observational tests (Sec. \ref{sec::flash}).  We analysed in-depth this simulation in both the Eulerian and Lagrangian frame, by employing our novel \CRaTer\ post-processing tracer code. In particular, we could isolate the key physical processes tied to the evolution of enstrophy/turbulence. Our main results are summarized as follows.
\begin{itemize}
 \item 
 The evolution of enstrophy follows the evolution of AGN activity, i.e. enstrophy is being amplified right after an AGN outburst (with a very short delay $\lta 2\ \Myr$). The amount of enstrophy amplification closely correlates with the mechanical power of the AGN (Fig. \ref{fig::evo_grid}). The regions of amplified enstrophy are significantly stronger along the jet cylinder rather than in the perpendicular/equatorial plane (Fig.~\ref{fig::eulerian_evo_shells}, \ref{fig::f_time_seq550_650}, \ref{fig::enstrophy_evo_4x2}). The enstrophy then starts to decay as turbulence propagates over the bulk of the ICM volume (Fig.~\ref{fig::rel_evo_grid}, \ref{fig::eulerian_evo_shells}, \ref{fig::f_time_seq550_650}) between lower levels of feeding CCA rain and AGN feedback injection. \\
 \item
 In the Eulerian (volume-wise) frame, the evolution of enstrophy depends, in principle, on its advection, compression, stretching, and baroclinic terms (Eq. 1 in the original paper). 
 However throughout most of the volume, the key drivers are found to be stretching and divergence/rarefaction motions, while advection becomes relevant only rarely and intermittently (Fig. 2, 3). Dominant vortex stretching (instead of squeezing) is a signature of subsonic/incompressible turbulence. 
 Baroclinicity is sub-dominant throughout largest part of the volume. However, it can become substantial in inner localized patches due to misaligned pressure and density gradients. We will perform a detailed analysis of the local properties of the baroclinic term in a forthcoming paper.\\
 \\
 \item
 In the Lagrangian (mass-wise) frame, tracer particles show that the AGN jet activity drives an envelope/cocoon of strong compression within which enstrophy is amplified (Fig. \ref{fig::fstr_evo_4x2}, \ref{fig::fcom_evo_4x2}). As ensemble, the growth of enstrophy is led by strong, localized stretching motions (Fig. \ref{fig::frel_evolution_all}). However, global rarefactions work against this runaway amplification of enstrophy, preserving a subsonic level of turbulence ($\rm Mach\sim 0.1-0.3$) over the cosmic time (e.g. Fig. \ref{fig::f_time_seq550_650}).
 Remarkably, positive compressions remain on average sub-dominant (besides the thin major-shock layer tied to the jet cocoon). This signals gentle self-regulated AGN feedback (Fig. \ref{fig::fcom_evo_4x2}). 
 For reference, enstrophy values typically oscillates in the range $10^{-3}-10^0\ \Myr^{-2}$.
 \\
 \item
 By selecting 16 families of tracer particle with different final positions (jets, lobes, and equatorial region), we find that the localized evolution of enstrophy and its dynamical terms are asymmetric with respect to the jet axis over short time periods.
 Regardless of the tracer location, each family evolution corroborates the Eulerian analysis that only stretching/rarefaction motions are the key for the growth/decline of enstrophy, while baroclinic motions remain sub-dominant (Sec. \ref{ssec::tracer_families}).
 Zooming into each macro family shows that turbulent motions can be significantly intermittent and asymmetric, even between the strongest southern/northern regions of the AGN injection. Most uptrends in enstrophy are tightly correlated with the preceding AGN outflow, bubble or cocoon injection, with largest/lowest scatter ($\sim3$/0.5 dex) occurring in the jets/equatorial tracer families, respectively.
 \\
 \item
 The above tracer ratios and trends are found to be valid during both high and low levels of AGN activity, with only different amplitudes. Thus, despite the  variations in the micro-scale kinematics, as space- and time-ensemble, the recurrent AGN feedback is able to sustain a background level of enstrophy/turbulence ($\sigma_v\sim 100\ \rm km\,s^{-1}$), superposed by recurrent spikes (up to $\sim 700\ \rm km\,s^{-1}$) due to impulsive anisotropic AGN features.
 Such duality is a common feature of chaotic systems, in which the micro evolution appears significantly random and intermittent, while more coherent and smooth properties emerge on coarser scales.
\end{itemize} 
Overall, the evolution of enstrophy and, hence, of turbulence is closely related to the activity of the AGN, and tends to be highly dynamic/chaotic over moderate intervals of space and time. Thereby, akin to Earth weather (\citealt{Bauer:2015}), when precise forecasts are needed, detailed hydrodynamical simulations with continuously updated boundary/internal conditions are required. Despite the complexity of such chaotic weather-like systems, we were able to isolate the dominant processes that lead to the amplification and dissipation of enstrophy: stretching motions amplify enstrophy, while rarefactions act as a counterbalancing sink. Despite the original jet/outflow being supersonic, we thus expect turbulence in hot halos to remain subsonic (${\rm Mach} < 1$) over several Gyr cycles of AGN feeding and feedback, as suggested by indirect X-ray and optical observations (e.g., \citealt{Hofmann:2016, Gaspari:2018, Simionescu:2019}). Remarkably, the enstrophy/turbulence decay and growth cycle mimics closely the AGN feeding and feedback cycle and self-regulation mechanism.\\
 \indent
Upcoming X-ray IFU telescopes, such as {\it Athena}\footnote{\url{https://www.the-athena-x-ray-observatory.eu}} (\citealt{Nandra:2013}) and {\it XRISM}\footnote{\url{https://heasarc.gsfc.nasa.gov/docs/xrism}} (\citealt{Kitayama:2014}) will soon resolve, pixel-by-pixel, the detailed X-ray spectral lines and thus thermo-kinematics of hot halos in nearby clusters and groups down to kpc scales, in particular via the broadening of ionized iron lines (e.g., \citealt{Ettori:2013, Cucchetti:2018, Roncarelli:2018}). The advancements and theoretical insights presented in this work (which is part of the broader \textit{BlackHoleWeather} program; \citealt{Gaspari:2020}) will be thus important to fully leverage the large spectral datasets delivered by such revolutionary X-IFU instruments, and to interpret the physical mechanisms and drivers behind the generation/evolution of turbulence in the diffuse gaseous halos of galaxies, groups, and clusters of galaxies.
 \section*{Acknowledgements}
 We thank the Reviewer for the fast feedback and constructive comments.
 D.W. is funded by the Deutsche Forschungsgemeinschaft (DFG, German Research Foundation) - 441694982.
 D.W. acknowledges former financial support from the European Union’s Horizon 2020 programme under the ERC Starting Grant ‘MAGCOW’, No. 714196.
 M.G. is supported by the \textit{Lyman Spitzer Jr.} Fellowship (Princeton University) and by NASA Chandra GO8-19104X/GO9-20114X and HST GO-15890.020-A grants.
 HPC resources were in part provided by the NASA/Pleiades HEC Program (SMD-7251).
 We are thankful for the `Multiphase AGN Feeding \& Feedback' workshop\footnote{\small \url{http://www.sexten-cfa.eu/event/multiphase-agn-feeding-feedback}} 
 (at SCfA, Italy)  which has stimulated helpful discussions. 
 We thank T. Jones, F. Vazza, and M. Br\"uggen for useful insights.
 
 \section*{Data availability}
 The data underlying this article will be shared on reasonable request to the authors, unless being in conflict or breaking the privacy of ongoing related work led by our collaboration members.

 \bibliographystyle{biblio.bst}
 \bibliography{mybib}
 \appendix
 \section{Complementary enstrophy and dynamical terms maps}\label{app::more_maps}
\begin{figure*}
     \centering
     \includegraphics[width = 0.98\textheight, angle = 90]{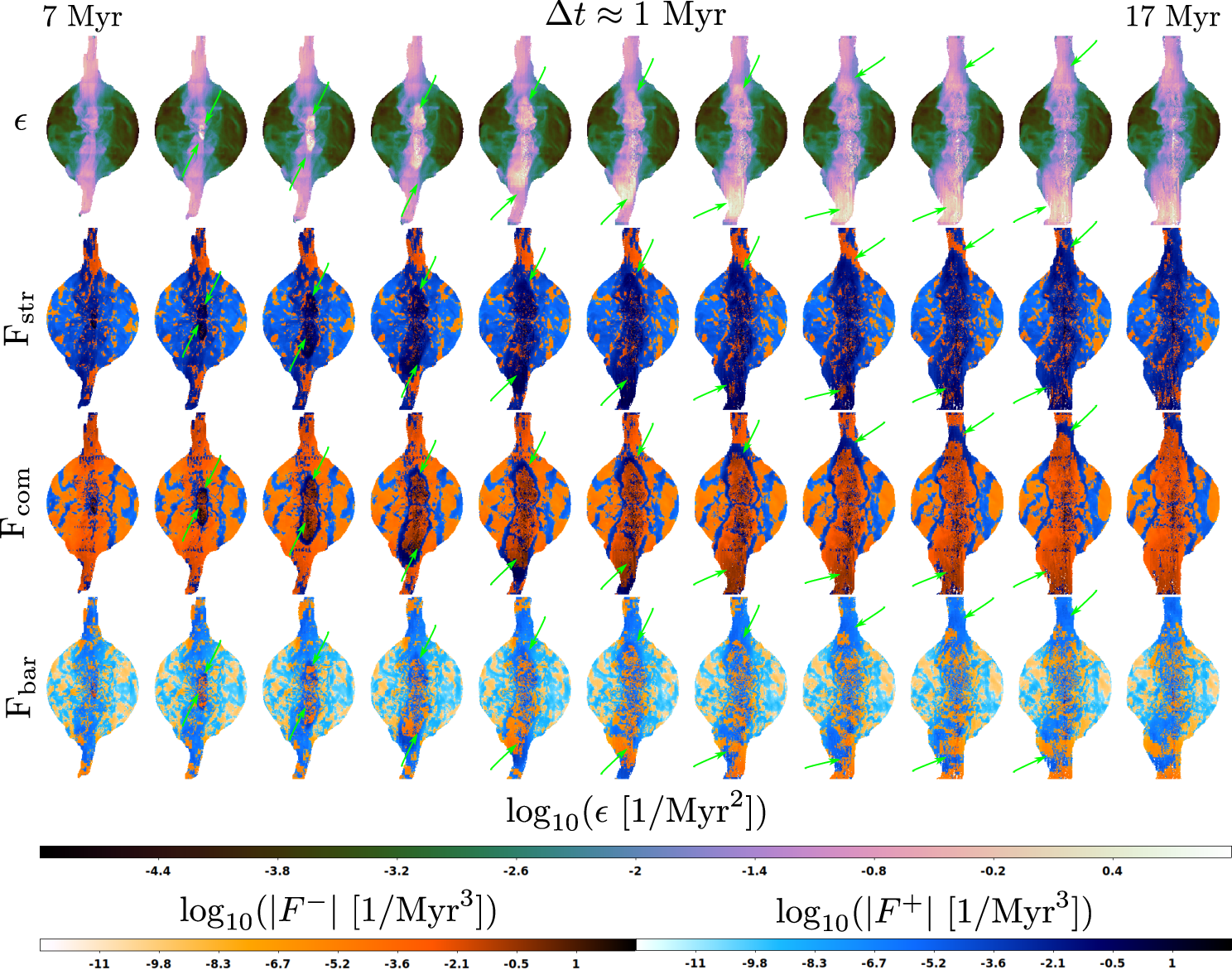} 
     \vspace{-0.5cm}
     \caption{Lagrangian analysis: projected enstrophy and dynamical terms (mass-weighted) recorded in the period $7-17 \ \Myr$ and displayed every $1\ \Myr$, which cover the first strong event of jet activity (Fig. \ref{fig::evo_grid}). The arrows mark regions of maximum enstrophy and are often correlated with the jet hot spots.}
     \label{fig::f_time_seq070_170}
 \end{figure*}

\begin{figure*}
     \centering
     \includegraphics[width = 0.98\textheight, angle = 90]{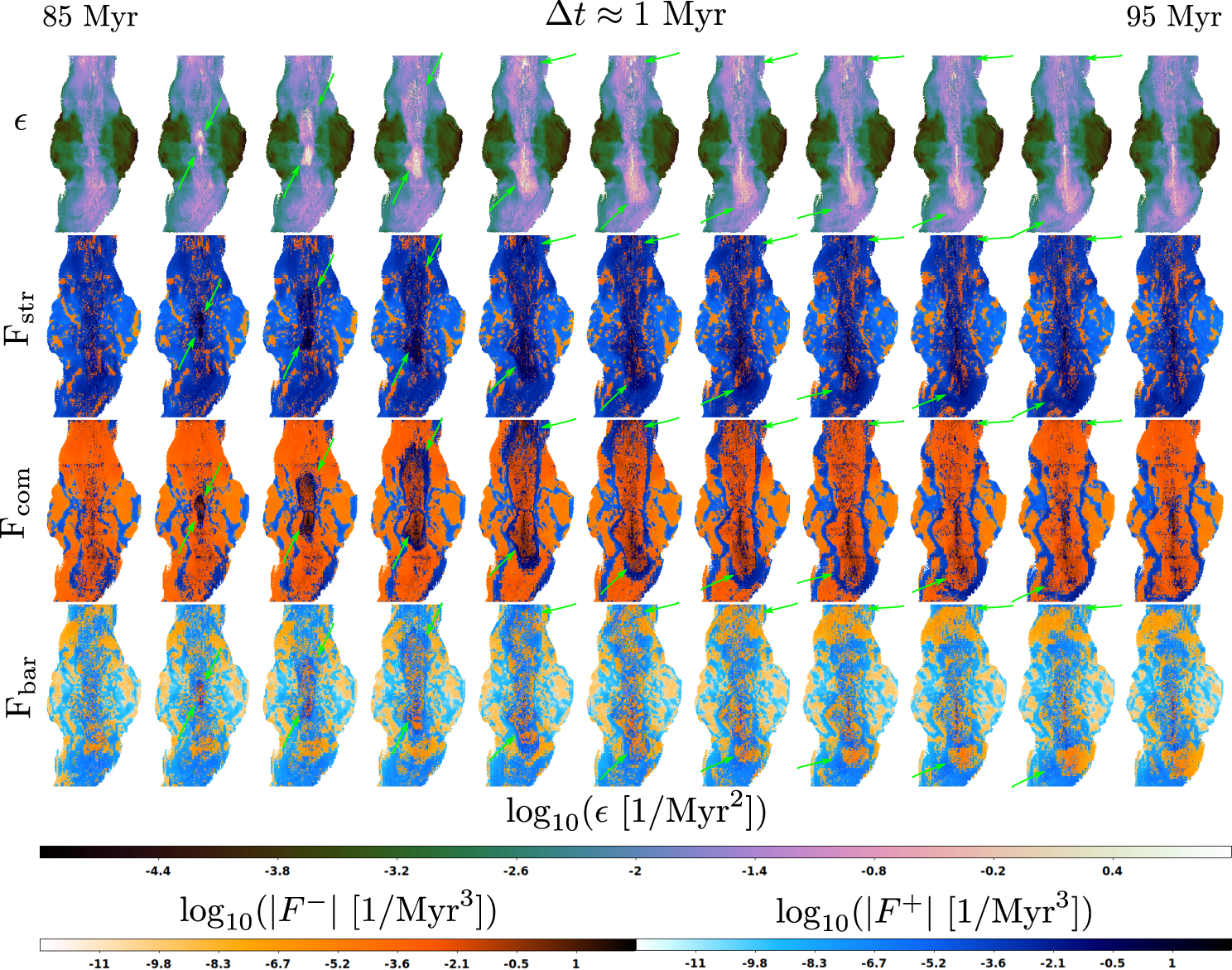} 
     \vspace{-0.5cm}     
     \caption{Lagrangian analysis: projected enstrophy and dynamical terms (analogue of Fig.~\ref{fig::f_time_seq070_170}) recorded in the period $85 \ \Myr$ to $95 \ \Myr$ covering the third strong event of jet activity (Fig. \ref{fig::evo_grid}).}
     \label{fig::f_time_seq850_950}
 \end{figure*}

 In the following, we show the maps of enstrophy and dynamical terms in the period $7-17 \ \Myr$ (Fig. \ref{fig::f_time_seq070_170}) and $85-95 \ \Myr$ (Fig. \ref{fig::f_time_seq850_950}). The formatting of the figures is the same as of Fig. \ref{fig::f_time_seq550_650} described in Sec.~\ref{sec::results_tracers}. The trends that were found and described in Sec. \ref{sec::results_tracers} are also found in Fig. \ref{fig::f_time_seq070_170} and \ref{fig::f_time_seq850_950}: enstrophy is mostly generated along the AGN feedback cocoon, where it propagates outwards. Stretching motions acting as a source term are the strongest along the jet axis. As counterbalance, compressions are mostly negative over the bulk of the ICM atmosphere, i.e. shaped by rarefactions. Yet, a thin envelope of strong compression surrounds the jet cocoon (shock), inside which enstrophy is also amplified. Baroclinicity acts both as a flickering sink and a source but remains always subdominant.\\
 \indent
 At the same time, it is important to note that the detailed features of single AGN event are highly variable, i.e., the outflow opening angle and precession, bubble diameter/elongation, and jet bipolarity, continuously change. The evolution of several AGN feeding and feedback cycles thus leads to the long-term enstrophy generation over the entire $4\pi$ solid angle, sustaining a stable average background of subsonic turbulence level over the cosmic time.

 \end{document}